\documentclass[12pt]{iopart}

\usepackage[colorlinks,linkcolor=black]{hyperref}
\usepackage{bm,graphicx,upgreek,amsmath,amssymb}
\usepackage{color,soul}
\newcommand{\rbd}{$^{87\!}{\rm Rb}$}
\newcommand*\colvec[3][]{\begin{pmatrix}\ifx\relax#1\relax\else#1\\\fi#2\\#3\end{pmatrix}}

\def\NT{\mathcal{N}_T}
\def\NL{\mathcal{N}_L}

\def\UpsilonT{\Upsilon_{\!T}}
\def\upsilonT{\upupsilon_{T}}

\begin{document}
	
\title[]{Mean-field dynamics of two-mode Bose-Einstein condensates in highly anisotropic potentials: Interference, dimensionality, and entanglement}

\author{Alexandre B.~Tacla}
\address{
	Center for Quantum Information and Control, MSC07--4220, University of New Mexico,
	Albuquerque, New Mexico 87131-0001, USA}
	\ead{tacla@unm.edu}
	
\author{Carlton M.~Caves}
\address{
	Center for Quantum Information and Control, MSC07--4220, University of New Mexico, Albuquerque, New Mexico 87131-0001, USA}
	\address{Centre for Engineered Quantum Systems, School of Mathematics and Physics, University of Queensland, Brisbane, Queensland 4072, Australia}

\begin{abstract}
We study the mean-field dynamics and the reduced-dimension character of two-mode Bose-Einstein condensates (BECs) in highly anisotropic traps.  By means of perturbative techniques, we show that the tightly confined (transverse) degrees of freedom can be decoupled from the dynamical equations at the expense of introducing additional effective three-body, attractive, intra- and inter-mode interactions into the dynamics of the loosely confined (longitudinal) degrees of freedom.  These effective interactions are mediated by changes in the transverse wave function.  The perturbation theory is valid as long as the nonlinear scattering energy is small compared to the transverse energy scales.  This approach leads to reduced-dimension mean-field equations that optimally describe the evolution of a two-mode condensate in general quasi-1D and quasi-2D geometries.  We use this model to investigate the relative phase and density dynamics of a two-mode, cigar-shaped $^{87}$Rb~BEC.  We study the relative-phase dynamics in the context of a nonlinear Ramsey interferometry scheme, which has recently been proposed as a novel platform for high-precision interferometry.  Numerical integration of the coupled, time-dependent, three-dimensional, two-mode Gross-Pitaevskii equations for various atom numbers shows that this model gives a considerably more refined analytical account of the mean-field evolution than an idealized quasi-1D description.
\end{abstract}


\maketitle	


\title[Mean-field dynamics of two-mode BECs in highly anisotropic potentials]{}

\vspace{2pc}

\section{Introduction	\label{sec:introduction}} 

Quantum protocols for nonlinear interferometry using two-mode Bose-Einstein condensates (BECs) have recently been proposed as a novel platform for weak-signal detection~\cite{rey07,boixo08b,choi08}.  More specifically, a two-mode $^{87}$Rb BEC consisting of $N$ atoms can be used to implement a nonlinear Ramsey interferometer that potentially operates near the limits established by quantum mechanics~\cite{boixo09,tacla10,taclathesis}.  This protocol is able to achieve detection sensitivities that scale better than the optimal $1/N$ limit of linear interferometry~\cite{giovannetti06a}, without relying on complicated state preparation or measurement procedures nor on entanglement generation to enhance the measurement sensitivity~\cite{boixo08b}.  For these reasons, this scheme can be particularly attractive from an experimental perspective.

As discussed in previous work~\cite{boixo09,tacla10,taclathesis}, there are several conditions for observing nonlinearity-enhanced scalings in this system.  For instance, the expansion of the condensate with increasing atom number can get in the way of achieving the desired scaling, as it essentially dilutes the nonlinear interactions in the condensate.  Although this effect can be completely suppressed by hard-walled homogeneous potentials, confining the condensate in highly anisotropic traps is a more realistic solution. In this way, the expansion of the atomic cloud in the tightly confined directions is effectively suppressed, with highly elongated geometries providing the architecture with best scalings.

In view of currently available techniques and typical experimental parameters, we numerically simulated such a nonlinear BEC interferometer in cigar-shaped (quasi-1D) potentials~\cite{tacla10}.  Interestingly, our simulations revealed that the interferometer's signal can be sensitive to the three-dimensional nature of the condensate.  In spite of the highly elongated confinement, we found significant deviations from a quasi-1D model as the strength of the nonlinear scattering interaction increases.  In this reduced-dimension approximation, it is assumed that the tightly confined (transverse) dimensions can be effectively neglected on the grounds that the characteristic transverse energy scale far exceeds the scattering interaction energy of the atomic cloud.  In this situation, however, there are still position-dependent phase shifts that need to be modeled precisely for an accurate analytical description of the interferometry process. This brings into question the accuracy of the reduced-dimension approximation, both spatially and temporally.

To better model the ground-state properties of a BEC in highly anisotropic potentials, we derived corrections to such approximation by means of perturbative techniques~\cite{tacla11}.  Using a perspective borrowed from quantum information theory, we developed a perturbative Schmidt decomposition of the condensate wave function between the transverse direction(s) and the loosely confined (longitudinal) direction(s).  This formalism provides corrections to the lowest-order transverse and longitudinal wave functions of the reduced-dimension approximation; the main effect is a reshaping of the BEC in the tightly confined direction(s) as the strength of the nonlinear scattering interaction increases.  In addition, because the perturbation formalism is tied to the Schmidt decomposition, it automatically encodes information about the entanglement between the spatial directions in higher-order Schmidt terms; the leading Schmidt term provides the optimal product approximation to the exact three-dimensional ground-state mean-field solution.

In this article, we develop further the approach of Ref.~\cite{tacla11} by extending our perturbation theory to the mean-field dynamics of two-mode BECs in highly anisotropic traps.  The perturbation theory is valid as long as the nonlinear scattering energy is small compared to the transverse energy scales.  This approach leads to equations that effectively describe the evolution of a two-mode condensate in general quasi-1D and quasi-2D geometries.  In addition, we also derive effective time-evolution equations for a highly anisotropic, single-mode BEC.  These equations show how the corrections to the reduced-dimension approximation propagate in time and affect the overall dynamics of the condensate, thus modeling not only interference effects, but also density dynamics.  We apply this 3D-corrected model to analyze the nonlinear BEC interferometry protocol simulated in Ref.~\cite{tacla10} for cigar-shaped potentials.  This analysis leads to a considerably more refined analytical account of the interference signal than the one given by a quasi-1D model.  In addition, we study the spatial segregation of the two modes, which occurs on a longer timescale than the phase accumulation.  We compare this model against exact three-dimensional numerical results for the evolution of the two-mode BEC given by the three-dimensional, two-mode, coupled Gross-Pitaevskii equations.  Finally, we investigate the reduced-dimension structure of the highly anisotropic, two-mode condensate by explicitly deriving the instantaneous Schmidt decomposition of the time-dependent, two-mode condensate state.  This analysis allows us to study the entanglement between the transverse and the longitudinal-internal degrees of freedom.  It confirms that the tightly confined dimensions can indeed be decoupled from the evolution equation and naturally shows the optimal way to do it.

This article is organized as follows.  In Sec.~\ref{sec:two_mode_dynamics_via_the_reduced_dimension_approximation}, we briefly recount the standard reduced-dimension approximation to the three-dimensional, two-mode, coupled Gross-Pitaevskii equations, which completely neglects the effect of the nonlinear scattering interaction on the transverse degrees of freedom.  By developing a perturbative relative-state decomposition of the time-dependent condensate mean fields, we derive corrections to the standard approximation in Sec.~\ref{sec:dynamics_via_the_relative_state_decomposition}; these corrections act as effective three-body, attractive, intra- and inter-mode interactions.  Because of the high anisotropy of the condensate, the evolution of the longitudinal modes in the perturbative regime takes place on a much longer time scale than the transverse modes.  We address this separation of time scales in \ref{sec:adiabatic_elimination_of_varphi_n} and derive an adiabatic approximation to the longitudinal evolution equations of Sec.~\ref{sec:dynamics_via_the_relative_state_decomposition}.  In Sec.~\ref{sec:two_mode_dynamics_of_a_cigar_shaped_87_rb_condensate}, we use the resulting model to investigate the relative-phase and density dynamics of a two-mode, cigar-shaped $^{87}$Rb BEC.  We study the relative-phase dynamics in the context of a nonlinear Ramsey interferometry scheme. We compare our perturbative model against numerical integration of the full three-dimensional, two-mode, coupled Gross-Pitaevskii equations for various atom numbers.  Lastly, in Sec.~\ref{sec:optimal_reduced_dimension_evolution_equations}, we investigate the reduced-dimension character of the two-mode condensate by explicitly deriving the instantaneous Schmidt decomposition of the time-dependent, two-mode condensate state.  Conclusions are given in Sec.\ref{sec:conclusion}.


\section{Two-mode dynamics in the standard reduced-dimension approximation} 
\label{sec:two_mode_dynamics_via_the_reduced_dimension_approximation}

In the mean-field approximation, one describes the evolution of a two-mode condensate at zero temperature by unity-normalized wave functions, $\psi^{1}(\mathbf r, t)$ and $\psi^{2}(\mathbf r, t)$, for the two modes, which are determined by the two-mode, coupled, time-dependent, three-dimensional Gross-Pitaevskii (GP) equations
\begin{align}
\label{twomodeGPE}
	i\hbar \frac{\partial \psi^{\alpha}}{\partial t\phantom{^\alpha}} = \Big( -\frac{\hbar^2}{2m}\nabla^2 + V + g_{\alpha\alpha}(N_{\alpha}-1)|\psi^{\alpha}|^2+ g_{\alpha\beta}N_{\beta}|\psi^{\beta}|^2 \Big)\psi^{\alpha}.
\end{align}
Here $N_{1(2)}$ is the number of atoms in mode 1(2), $V$ is the external trapping potential, assumed to be the same for both modes, and $\alpha=1,2$ and $\beta=2,1$ label the two modes of the condensate, which can represent, for instance, two hyperfine states, $|1\rangle$ and $|2\rangle$.  We further assume that collisions between the atoms are elastic, so that the only allowed scattering processes are $|1\rangle | 1
\rangle \rightarrow |1\rangle | 1 \rangle$, $|2\rangle | 2
\rangle \rightarrow |2\rangle | 2 \rangle$, and $|1\rangle | 2
\rangle \rightarrow |1\rangle | 2 \rangle$, with scattering
strengths $g_{11}$, $g_{22}$, and $g_{12}$, where
$g_{\alpha\beta} = 4 \pi \hbar^2
a_{\alpha\beta}/m$ is determined by the $s$-wave scattering length $a_{\alpha \beta}$ and the atomic mass $m$.  Note that by setting the inter-mode coupling constant $g_{12}$ to zero, one recovers the case of a single-mode BEC, which we address in more detail in \ref{sub:single_mode_dynamics_within_the_adiabatic_elimination_of_upvarphi_n_alpha}.

In the case of highly anisotropic potentials, the condensate is loosely trapped by a potential $V_L({\bm r})$ in $d$ dimensions, referred to as longitudinal ($L$) dimensions, as opposed to the remaining $D=3-d$ transverse degrees of freedom ($T$), which are tightly confined in a potential $V_T({\bm\rho})$. If the scattering interaction is sufficiently small compared to the transverse energy scale, one can neglect the effect of the nonlinear interaction on the atomic transverse degrees of freedom and approximate the condensate wave functions by the product ansatz
\begin{equation}
\label{ansatz}
    \uppsi_{\rm rda}^{\alpha}({\bm\rho},{\bm r},t) = e^{-i E_0 t/\hbar}\xi_0({\bm \rho})\phi^{\alpha}({\bm r},t),
\end{equation}
where $E_0$ and $\xi_0({\bm\rho})$ are, respectively, the ground-state energy and wave function of the bare transverse potential, $V_T({\bm\rho})$.  The longitudinal wave functions $\phi^{\alpha}({\bm r},t)$ are the solutions of the $d$-dimensional, longitudinal GP equations,
\begin{equation}
\label{dDGPE}
i\hbar \frac{\partial \phi^{\alpha}}{\partial t\phantom{^\alpha}}
=\left(-\frac{\hbar^{2}}{2 m}\nabla_L^2+V_L + g_{\alpha\alpha}\eta_T(N_{\alpha}-1)|\phi^{\alpha}|^2+ g_{\alpha\beta}\eta_T N_{\beta}|\phi^{\beta}|^2 \right)\!\phi^\alpha,
\end{equation}
which are found by plugging Eq.~(\ref{ansatz}) into the two-mode GP equations~(\ref{twomodeGPE}) and projecting the result onto the subspace spanned by $\xi_0$.  Note that in this standard reduced-dimension picture, the coupling constants are renormalized by the average inverse transverse cross section of the condensate, given by
\begin{equation}
\eta_T=\int d^D\!\rho\,|\xi_0({\bm \rho})|^4.
\end{equation}
Within this approximation, the transverse and longitudinal degrees of freedom are decoupled.

This approximation is only meaningful if the number of atoms in the condensate is small compared to an (upper) critical atom number $\NT$, defined as the number of atoms at which the nonlinear scattering energy becomes comparable to the transverse kinetic energy.  For a \rbd\ condensate, with the experimentally accessible trap parameters we consider in detail in Sec.~\ref{sec:two_mode_dynamics_of_a_cigar_shaped_87_rb_condensate}, $\NT$ has the value $14\,000$.  As $N$ approaches $\NT$, one can no longer neglect the effects of the scattering interaction on the condensate transverse degrees of freedom; as a result, the product ansatz~(\ref{ansatz}) is no longer a good approximation to the 3D wave function. Such 3D-induced effects are responsible not only for modifying the transverse and longitudinal wave functions, but also for creating correlations between the spatial directions.  All these effects can be readily calculated in the perturbative regime where $N$ is small compared to $\NT$.


\section{Two-mode dynamics via the relative-state decomposition} 
\label{sec:dynamics_via_the_relative_state_decomposition}

In previous work~\cite{tacla11}, we found perturbative corrections to the standard reduced-dimension approximation for the ground-state wave function of a single-mode condensate.  These corrections arise from the derivation of the Schmidt decomposition of the condensate wave function between the transverse and longitudinal degrees of freedom.  To first-order this approach is equivalent to a first-order perturbative relative-state decomposition of the condensate wave function.  The crucial difference between these two approaches is that the Schmidt decomposition assumes no prior knowledge of the basis elements used in the perturbation expansion, whereas in the relative-state method, the expansion is carried out relative to a fixed basis for the transverse degrees of freedom.  For this reason, the expansion of the time-independent condensate wave function can be implemented in a simpler way if derived via the relative-state method rather than via the Schmidt decomposition.  This motivates us to investigate the dynamics of the condensate from a similar perspective and to look for the perturbative relative-state decomposition of the time-dependent condensate mean field.  We return to the Schmidt decomposition in Sec.~\ref{sec:optimal_reduced_dimension_evolution_equations} and show that a Schmidt decomposition of the three-dimensional, time-dependent mean-field solution can easily be retrieved from the relative-state decomposition. The Schmidt decomposition fully characterizes the entanglement among the condensate's spatial co\"ordinates and its internal degrees of freedom and thus provides an optimal method for investigating the reduced-dimension character of the condensate.

As before, we model the two-mode dynamics of the condensate by means of the mean-field approximation, according to which the mean field of each BEC mode satisfies the time-dependent, coupled, two-mode Gross-Pitaevskii equation~(\ref{twomodeGPE}), which we write here as
\begin{align}
\label{perturbtwomodeGPE}
	i\hbar \dot{\psi}^{\alpha} = \Big( H_{T} + \epsilon H_{L} + \epsilon \tilde{g}_{\alpha\alpha}|\psi^{\alpha}|^2+ \epsilon\tilde{g}_{\alpha\beta}|\psi^{\beta}|^2 \Big)\psi^{\alpha}.
\end{align}
For simplicity of notation, we now use $\dot{\psi}^{\alpha}=\partial\psi^{\alpha}/\partial t$, $\tilde g_{\alpha\alpha}=(N_{\alpha}-1)g_{\alpha\alpha}$, and $\tilde{g}_{\alpha \beta} = N_{\beta} g_{\alpha \beta}$.  Here $H_{T(L)}=-(\hbar^2/2m)\nabla_{T(L)}^2 + V_{T(L)}$ is the transverse (longitudinal) single-particle Hamiltonian, and $\epsilon$ is a formal perturbation parameter that is set equal to 1 at the end of the calculation.  We discuss the physical dimensionless expansion parameter for the perturbation theory in Sec.~\ref{sub:numerical_results}.

In the perturbative regime, we write the relative-state decomposition of the time-dependent condensate wave functions $\psi^{\alpha}$ as
\begin{align}
\label{timedependentpsiRSD}
\psi^{\alpha}({\bm \rho},{\bm r},t)= \xi_0({\bm \rho})\varphi^{\alpha}_0({\bm r},t)+
\epsilon\sum_{n=1}^\infty \xi_n({\bm \rho})\varphi^{\alpha}_n({\bm r},t),
\end{align}
where $\{\xi_n\}$ is the eigenbasis of the transverse Hamiltonian.  The longitudinal wave functions $\varphi^{\alpha}_n$ are defined by the projection of the time-dependent mean-field solution $\psi^{\alpha}$ onto the transverse eigenfunctions $\xi_n$.

Before we carry out the perturbative expansion of the time-dependent, two-mode GP equations, it is convenient to redefine the longitudinal wave functions relative to an interaction picture in which fast, trivial oscillations are removed.  Because of the high anisotropy of the trapping potential, we expect the transverse bare trap energy to be the fastest time scale in the perturbative regime, and therefore we define
\begin{align}
	\upvarphi^{\alpha}_n(t) \equiv  e^{i E_0 t/\hbar} \varphi^{\alpha}_n(t).
\end{align}

Now we expand the two-mode GP equations~(\ref{perturbtwomodeGPE}) to second order in powers of $\epsilon$ and project the result onto $\xi_0$, thus obtaining the following equation for $\upvarphi^{\alpha}_0$:
\begin{align}
\label{vphi0expansion}
	i\hbar \dot{\upvarphi}^{\alpha}_{0} = \epsilon &\left( H_L + \tilde{g}_{\alpha \alpha} \eta_T |\upvarphi^{\alpha}_{0}|^{2} + \tilde{g}_{\alpha \beta} \eta_T |\upvarphi^{\beta}_{0}|^{2} \right)\upvarphi^{\alpha}_{0} \nonumber\\
	& + \epsilon^{2} \sum_{n=1}^{\infty} \langle \xi_{0}^{3} | \xi_{n} \rangle \Big[  \left( 2\tilde{g}_{\alpha \alpha} |\upvarphi_0^{\alpha}|^{2} + \tilde{g}_{\alpha \beta} |\upvarphi_0^{\beta}|^{2} \right) \upvarphi_n^{\alpha} +  \tilde{g}_{\alpha \alpha} (\upvarphi_0^{\alpha})^{2} \upvarphi_n^{\alpha*}\Big] \nonumber\\
	& + \epsilon^{2}  \tilde{g}_{\alpha \beta} \upvarphi_0^{\alpha} \sum_{n=1}^{\infty} \langle \xi_{0}^{3} | \xi_{n} \rangle \left( \upvarphi_0^{\beta}\upvarphi_n^{\beta*} + \upvarphi_0^{\beta*}\upvarphi_n^{\beta} \right).
\end{align}
Hereafter, for brevity, we represent spatial integrals in terms of bra-ket inner products.

Similarly, the projection of Eq.~(\ref{perturbtwomodeGPE}) onto $\xi_n$, $n\ge1$, yields to lowest order in $\epsilon$ the time-evolution equations for the longitudinal functions $\upvarphi^{\alpha}_n$, which read
\begin{align}
\label{vphindot}
	i\hbar\dot{\upvarphi}_n^{\alpha} = (E_n - E_0) \upvarphi^{\alpha}_{n} + \langle \xi_{n} | \xi_{0}^{3} \rangle \left( \tilde{g}_{\alpha \alpha} |\upvarphi^{\alpha}_{0}|^{2} + \tilde{g}_{\alpha \beta} |\upvarphi^{\beta}_{0}|^{2} \right) \upvarphi^{\alpha}_{0} + O(\epsilon),
\quad n\ge 1.
\end{align}
One can integrate Eq.~(\ref{vphindot}), obtaining the formal solution
\begin{align}
\label{vphinsol}
	\upvarphi_n^{\alpha}(t) &- \upvarphi_n^{\alpha}(0) e^{-i (E_n-E_0)t/\hbar}\nonumber\\
&= -\frac{i}{\hbar}\langle \xi_{n} | \xi_{0}^{3} \rangle \int_{0}^{t} ds\, e^{-i (E_n-E_0)(t-s)/\hbar}  \Big( \tilde{g}_{\alpha \alpha} |\upvarphi^{\alpha}_{0}(s)|^{2}
+ \tilde{g}_{\alpha \beta} |\upvarphi^{\beta}_{0}(s)|^{2} \Big) \upvarphi^{\alpha}_{0}(s) + O(\epsilon).
\end{align}

By plugging Eq.~(\ref{vphinsol}) back into Eq.~(\ref{vphi0expansion}), we eliminate $\upvarphi_n^{\alpha}(t)$ from the equation and get
\begin{align}
\label{vphi0}
	i\hbar \dot{\upvarphi}^{\alpha}_{0} = \epsilon &\left( H_L + \tilde{g}_{\alpha \alpha} \eta_T |\upvarphi^{\alpha}_{0}|^{2} + \tilde{g}_{\alpha \beta} \eta_T |\upvarphi^{\beta}_{0}|^{2} \right)\upvarphi^{\alpha}_{0}\nonumber\\
&+ \epsilon^{2} \eta_T^2 \left( 2\tilde{g}_{\alpha \alpha} |\upvarphi_0^{\alpha}|^{2} + \tilde{g}_{\alpha \beta} |\upvarphi_0^{\beta}|^{2} \right)\nonumber\\
	&\qquad\times\int_{0}^{t} ds \left( \tilde{g}_{\alpha \alpha} |\upvarphi^{\alpha}_{0}(s)|^{2} + \tilde{g}_{\alpha \beta} |\upvarphi^{\beta}_{0}(s)|^{2} \right) G_T(t-s) \upvarphi^{\alpha}_{0}(s)\nonumber\\
&+ \epsilon^{2}\eta_T^2\tilde{g}_{\alpha \alpha} (\upvarphi_0^{\alpha})^{2} \int_{0}^{t} ds \left( \tilde{g}_{\alpha \alpha} |\upvarphi^{\alpha}_{0}(s)|^{2} + \tilde{g}_{\alpha \beta} |\upvarphi^{\beta}_{0}(s)|^{2} \right) G_T^{*}(t-s)\upvarphi^{\alpha*}_{0}(s) \nonumber\\
	& + \epsilon^{2}\eta_T^2  \tilde{g}_{\alpha \beta} \upvarphi_0^{\alpha} \int_{0}^{t} ds \left( \tilde{g}_{\beta \beta} |\upvarphi^{\beta}_{0}(s)|^{2} + \tilde{g}_{\beta \alpha} |\upvarphi^{\alpha}_{0}(s)|^{2} \right) \Big( \upvarphi_0^{\beta} G_T^{*}(t-s)\upvarphi^{\beta*}_{0}(s) \nonumber\\
	&\qquad + \upvarphi_0^{\beta*} G_T(t-s) \upvarphi^{\beta}_{0}(s) \Big)\nonumber\\
&+ \epsilon^{2}  I^{\alpha}_0(t).
\end{align}
Here
\begin{align}
\label{GT}
	G_T(t) \equiv -\frac{i}{\hbar}\sum_{n=1}^{\infty} \frac{\langle \xi_{0}^{3} | \xi_{n} \rangle^2}{\eta_T^2} e^{-i (E_n-E_0)t/\hbar}
\end{align}
is a temporal response function that comes from changes in the transverse wave function
and that oscillates at the bare transverse eigenfrequencies, and
\begin{align}
	I^{\alpha}_0(t) = &\left( 2\tilde{g}_{\alpha \alpha} |\upvarphi_0^{\alpha}|^{2} + \tilde{g}_{\alpha \beta} |\upvarphi_0^{\beta}|^{2} \right) \sum_{n=1}^{\infty} \langle \xi_{0}^{3} | \xi_{n} \rangle   \upvarphi_n^{\alpha}(0) e^{-i (E_n-E_0)t/\hbar} \nonumber\\
	& + \tilde{g}_{\alpha \alpha} (\upvarphi_0^{\alpha})^{2} \sum_{n=1}^{\infty} \langle \xi_{0}^{3} | \xi_{n} \rangle \upvarphi_n^{\alpha*}(0) e^{i (E_n-E_0)t/\hbar} \nonumber\\
	& + \tilde{g}_{\alpha \beta} \upvarphi_0^{\alpha} \sum_{n=1}^{\infty} \langle \xi_{0}^{3} | \xi_{n} \rangle \left( \upvarphi_0^{\beta}\upvarphi_n^{\beta*}(0) e^{i (E_n-E_0)t/\hbar} + \upvarphi_0^{\beta*}\upvarphi_n^{\beta}(0) e^{-i (E_n-E_0)t/\hbar} \right)
\end{align}
is the term associated with the initial condensate wave function.

Equations~(\ref{vphinsol}) and (\ref{vphi0}) are the results of the relative-state method.  Not surprisingly, $\upvarphi^{\alpha}_{0}$ satisfies to lowest order the standard reduced-dimension description of the time-dependent GP equations discussed in Sec.~\ref{sec:two_mode_dynamics_via_the_reduced_dimension_approximation}.  The higher-order terms, on the other hand, introduce corrections to the standard reduced-dimension approximation that act as effective three-body, attractive, intra- and inter-mode interactions.  These interactions are mediated by changes in the transverse wave function and take place on the transverse time scale set by the temporal response function $G_T$.

\begin{figure}[htbp]
	\centering
		\includegraphics[width=\textwidth]{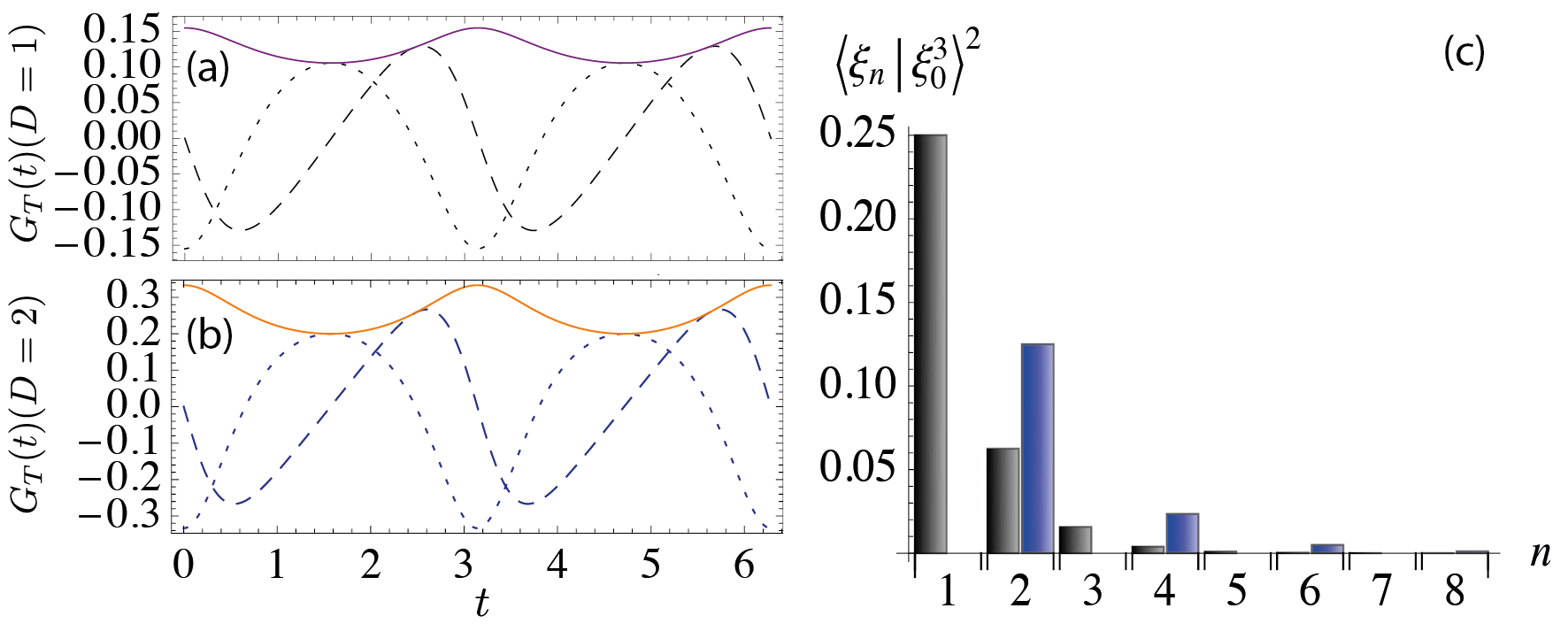}
	\caption{Temporal response function $G_T$, in units of $\hbar^{-1}$, for a transverse harmonic potential and (a) a pancake-like trap ($D=1$) and (b) a cigar-shaped trap ($D=2$). Time is measured in units of $\omega_T^{-1}$.  The dashed line represents the real part of $G_T$, the dotted line is the imaginary part, and the solid line is the absolute value of $G_T$.  $G_T$ is a periodic function with fundamental oscillation frequency equal to $2\omega_T$.  The anharmonicity of $G_T$ essentially comes from the lowest two nonvanishing modes in the spectral decomposition~(\ref{GT}), as shown in (c), where we plot the value of the coefficients $\langle \xi_{0}^{3} | \xi_{n} \rangle^2$ (in units of $\eta_T^2$) for a harmonic transverse trap for the first eight terms in the series~(\ref{GT}). For every $n$, we show the value of the coefficient for a cigar-shaped potential (black bar on the left) as well as for a pancake-like trap  (blue bar on the right).  For a quasi-1D potential ($D=2$), $n$ represents the radial quantum number of the harmonic potential.  For both cases, the spectral representation of $G_T$ reveals a single dominant mode, namely $n=1$ for a cigar and $n=2$ for a pancake-like potential (for $D=1$ the well-defined parity of the transverse ground state requires $\xi_{0}^3$ to be orthogonal to states with odd parity), that oscillates with frequency equal to $2\omega_T$.  The remaining, smaller terms introduce the anharmonicity seen in (a) and (b).}
	\label{fig:GT}
\end{figure}

Interestingly, the temporal response function~(\ref{GT}) has a closed-form solution for the case of a transverse harmonic potential.  In this case, the transverse ground-state wave function is the Gaussian
\begin{equation}
\label{gaussian}
	\xi_{0}({\bm \rho}) = \frac{e^{-\rho^2/2\rho_0^2}}{(\pi\rho_0^2)^{D/4}},
\end{equation}
where $\rho_0 = \sqrt{\hbar/m\omega_T}$, with $\omega_T$ being the transverse trap frequency.  It is easy to see that
\begin{equation}
\label{etaTD}
	\eta_{T} = \left(\frac{1}{\sqrt{2\pi}\rho_0}\right)^{D}.
\end{equation}
Moreover, for a pancake-like potential ($D=1$), we have
\begin{equation}\label{overlap1D}
\langle \xi_{n} | \xi_0^3 \rangle=
\begin{cases}
\displaystyle{\frac{(-1)^{n/2}}{\sqrt{\pi\, n!}}\eta_T \Gamma\left(\frac{n + 1}{2}\right)},
&\mbox{$n$ even,}\\
0,&\mbox{$n$ odd.}
\end{cases}
\end{equation}
For a cigar ($D=2$), if we use polar co\"ordinates for the transverse eigenfunctions, they take the form $\xi_{n_r m}(\rho,\varphi)$, with $n_{r}$ and $m$ being radial and azimuthal quantum numbers and with the eigenenergies given by $E_{n_r m}=\hbar\omega_T(2n_r+|m|+1)$.  Then we find that
\begin{equation}
\langle \xi_{n_{r}m} | \xi_{00}^3 \rangle=2^{-n_{r}}\eta_T\delta_{m0}.\label{overlap2D}
\end{equation}
By using Eqs.~(\ref{overlap1D}) and (\ref{overlap2D}), we can write the response function as
\begin{align}
\label{GTharmonic}
G_T(t)=
\begin{cases}
\displaystyle{\vphantom{\Bigg(\biggr)}\frac{i}{\hbar}\left(1-\frac{2}{\sqrt{4-e^{-2 i\omega_T t}}}\right) ,}&\mbox{$D=1$ (pancake),}\\
\displaystyle{\vphantom{\Bigg(\biggr)}\frac{i}{\hbar}\frac{1}{1-4 e^{2i\omega_T t} },}&\mbox{$D = 2$ (cigar).}
\end{cases}
\end{align}
Figure~\ref{fig:GT} shows the temporal response function~(\ref{GTharmonic}) for (a) pancake- and (b) cigar-shaped traps.

Due to the high anisotropy of the trapping potential, we expect the evolution of the longitudinal modes to take place on a much longer time scale than the transverse modes and the temporal response function $G_T$.  In fact, our perturbation expansion is valid as long as this assumption holds, for according to Eq.~(\ref{vphi0}), $\dot{\upvarphi}^{\alpha}_0 = O(\epsilon)$.  We can use this fact to derive an adiabatic approximation for $\upvarphi^{\alpha}_n$, which we then use to bring Eq.~(\ref{vphi0}) to a more tractable form.  We relegate the details of this derivation to \ref{sec:adiabatic_elimination_of_varphi_n}.  In the next section, we use the results, given in Eqs.~(\ref{vphinapprox}) and (\ref{vphi0final}), to investigate the relative-phase and density dynamics of a two-mode, cigar-shaped \rbd\ BEC. In \ref{sub:single_mode_dynamics_within_the_adiabatic_elimination_of_upvarphi_n_alpha} we give the dynamical equations for a single-mode BEC within the adiabatic approximation to $\upvarphi^{\alpha}_n$.


\section{Two-mode dynamics of a cigar-shaped $^{87}$Rb condensate} 
\label{sec:two_mode_dynamics_of_a_cigar_shaped_87_rb_condensate}

In this section we apply the general results of the time-dependent relative-state decomposition to analyze the two-mode dynamics of a cigar-shaped $^{87}$Rb condensate consisting of $N$ atoms that can occupy the $5S_{1/2}$ $|F=1;\,M_F=-1\rangle$ and $|F=2;\,M_F=+1\rangle$ hyperfine states.  These states are of particular interest to us because they can be used to implement a nonlinear Ramsey interferometry protocol with sensitivity that scales better than $1/N$, as discussed previously in Refs.~\cite{boixo08b,boixo09,tacla10,taclathesis}.

\subsection{Two-mode evolution} 
\label{sub:two_mode_evolution}

In typical experiments~\cite{matthews98a,hall98a,hall98b,mertes07a, anderson09, egorov11,egorov12}, the $|F=1;\,M_F=-1\rangle\equiv|1\rangle$ state is
trapped and cooled to the condensation point.  Once the atoms in
$|1\rangle$ have accumulated in the condensate ground state, a
two-photon drive is used to couple the $|1\rangle$ state to the
$|F=2;\,M_F=+1\rangle\equiv|2\rangle$ state.  These two states have nearly identical magnetic moments and hence feel essentially the same confining potentials.  This strategy is used instead of cooling the two hyperfine states simultaneously to form a condensate in a superposition, because the lifetime of atoms in the $|2\rangle$ state in a trap is much shorter than the lifetime of atoms in the $|1\rangle$ state.  Thus, we assume from now on that the optical pulse suddenly creates the superposition state $\psi({\mathbf r})(|1\rangle +
|2\rangle)/\sqrt{2}$ for each atom~\cite{note1}.  This procedure creates the two-mode condensate with both modes being occupied by the same number of atoms and having the same initial wave function $\psi({\mathbf r})$.  In this case, we have $\tilde g_{11}=g_{11}(N/2-1)$ and $\tilde g_{22}=g_{22}(N/2-1)$.  Moreover, the inter-mode coupling is symmetric, i.e.,
\begin{equation}
\tilde{g}_{12} = \tilde{g}_{21} = g_{12}N/2,
\end{equation}
which allows us to write the longitudinal two-mode equations~(\ref{vphi0final}) as
\begin{align}
\label{vphi01}
	i\hbar \dot{\upvarphi}^{1}_{0}
	&=\epsilon\left( H_L + \tilde{g}_{11} \eta_T |\upvarphi^{1}_{0}|^{2} + \tilde{g}_{1 2} \eta_T |\upvarphi^{2}_{0}|^{2} \right)\upvarphi^{1}_{0} \nonumber\\
	& \quad -\epsilon^{2} \UpsilonT \Big( 3 \tilde{g}_{1 1}^2 |\upvarphi^{1}_{0}|^{4} + (4 \tilde{g}_{1 1} + 2\tilde{g}_{12})\tilde{g}_{1 2} |\upvarphi_0^{1}|^{2}|\upvarphi^{2}_{0}|^{2}
+ \tilde{g}_{1 2}(\tilde{g}_{1 2} + 2\tilde{g}_{2 2}) |\upvarphi_0^{2}|^{4}\Big)\upvarphi^{1}_{0}+ \epsilon^{2}  \tilde{I}^{1}_0,\\
	i\hbar \dot{\upvarphi}^{2}_{0}
	&=\epsilon\left( H_L + \tilde{g}_{2 2} \eta_T |\upvarphi^{2}_{0}|^{2} + \tilde{g}_{12} \eta_T |\upvarphi^{1}_{0}|^{2} \right)\upvarphi^{2}_{0} \nonumber\\
	&\quad - \epsilon^{2} \UpsilonT \Big( 3 \tilde{g}_{2 2}^2 |\upvarphi^{2}_{0}|^{4} + (4 \tilde{g}_{2 2} + 2\tilde{g}_{1 2})\tilde{g}_{1 2} |\upvarphi_0^{2}|^{2}|\upvarphi^{1}_{0}|^{2}
+ \tilde{g}_{12}(\tilde{g}_{12} + 2\tilde{g}_{1 1}) |\upvarphi_0^{1}|^{4}\Big)\upvarphi^{2}_{0}+ \epsilon^{2}  \tilde{I}^{2}_0.\label{vphi02}
\end{align}
Here $\tilde{I}^{1}_0$ and $\tilde{I}^{2}_0$, defined by Eq.~(\ref{I0alpha}), are source terms associated with the initial condensate wave functions.  The coupling parameter
\begin{equation}\label{UpsilonT}
\UpsilonT\equiv
\sum_{n=1}^\infty\frac{\langle\xi_n|\xi_0^3\rangle^2}{E_n-E_0}\ge0
\end{equation}
is determined solely by the properties of the transverse trap and characterizes the strength of the coupling of the transverse and longitudinal directions.  The quantity $\eta_T^2/\UpsilonT$ can be thought of as the relevant quantification of the transverse energy scale as far as the perturbation theory is concerned.

Equations~(\ref{vphi01}) and (\ref{vphi02}) are written under the assumption that the longitudinal functions $\upvarphi^{1}_{n}(t)$ and $\upvarphi^{2}_{n}(t)$ can be adiabatically approximated according to Eq.~(\ref{vphinapprox}) and hence are given by
\begin{align}
\label{vphin1}
\upvarphi_n^{1}(t) &= - \frac{\langle \xi_{n} | \xi_{0}^{3} \rangle}{E_n-E_0} \left( \tilde{g}_{1 1} |\upvarphi^{1}_{0}(t)|^{2} + \tilde{g}_{12} |\upvarphi^{2}_{0}(t)|^{2} \right) \upvarphi^{1}_{0}(t) + \tilde{I}_n^{1}(t), \\
\upvarphi_n^{2}(t) &= - \frac{\langle \xi_{n} | \xi_{0}^{3} \rangle}{E_n-E_0} \left( \tilde{g}_{22} |\upvarphi^{2}_{0}(t)|^{2} + \tilde{g}_{12} |\upvarphi^{1}_{0}(t)|^{2} \right) \upvarphi^{2}_{0}(t) + \tilde{I}_n^{2}(t)\label{vphin2}.
\end{align}
Here $\tilde{I}^{1}_n$ and $\tilde{I}^{2}_n$, defined by Eq.~(\ref{Inalpha}), are source terms associated with the initial condensate wave functions. The remaining term, $\tilde\upvarphi_n^\alpha\equiv\upvarphi_n^\alpha - \tilde{I}^{\alpha}_n$, defined in Eq.~(\ref{vphinadiabatic}), corresponds to the adiabatic following of $|\upvarphi^{\alpha}_{0}|^{2}\upvarphi^{\alpha}_{0}$ and $|\upvarphi^{\beta}_{0}|^{2}\upvarphi^{\alpha}_{0}$ and is the dominant dynamical effect described by Eqs.~(\ref{vphin1}) and (\ref{vphin2}).  We use below the results from the relative-state decomposition of the ground-state wave function, derived in Ref.~\cite{tacla11}, to bring all the source terms to a simpler form.

In our mean-field model, we suppose that all $N$ atoms initially occupy the same single-particle wave function $\psi({\bm \rho},{\bm r})$, which is the solution of the time-independent, three-dimensional GP equation for the state $|1\rangle$.  For this particular choice of the initial state, we showed in Ref.~\cite{tacla11} that, in the perturbative regime $N\ll \NT$, the condensate wave function can be approximated, to first order in perturbation theory, by the time-independent relative-state decomposition
\begin{align}
	\psi({\bm \rho},{\bm r}) &= \xi_0({\bm \rho}) \varphi_{0}({\bm r}) + \epsilon \sum_{n=1}^{\infty} \xi_{n}({\bm \rho}) \varphi_{n}({\bm r})
	= \psi^{1}(t=0) = \psi^{2}(t=0),
\end{align}
where the dominant longitudinal wave function $\varphi_{0}({\bm r})$ is the solution to
\begin{equation}
\label{quinticequation}
 (\tilde{\mu}_L - H_L - \tilde{g} \eta_{T} \varphi_0^2 + 3\epsilon \tilde{g}^2 \UpsilonT \varphi_0^4) \varphi_0 = 0,
\end{equation}
with $\tilde{g}=(N-1)g_{11}$, and
\begin{align}
	\varphi_{n0} &= -\tilde{g}\varphi_{00}^3\frac{\langle \xi_{0}^{3} | \xi_{n} \rangle}{E_n-E_0}= \upvarphi^{1}_{n}(t=0) = \upvarphi^{2}_{n}(t=0).\label{vphinr0}
\end{align}
Here $\varphi_{00}$ is the zero-order longitudinal wave function given by the reduced-dimension, single-mode GP equation
\begin{equation}
\label{vphi00}
(\mu_L - H_L - \tilde{g} \eta_{T} \varphi_{00}^2) \varphi_{00} = 0.
\end{equation}

Thus, using Eq.~(\ref{vphinr0}), we can write Eq.~(\ref{Inalpha}) as
\begin{equation}
	\tilde{I}_n^{\alpha}(t) = -\big( \tilde{g} - \tilde{g}_{\alpha \alpha} -\tilde{g}_{\alpha \beta}\big) \varphi_{00}^3\frac{\langle \xi_{0}^{3} | \xi_{n} \rangle}{E_n-E_0} e^{-i (E_n-E_0)t/\hbar}\label{Inalphaapprox}
\end{equation}
and the functions~(\ref{Phia}) and (\ref{Phiab}) as
\begin{align}
	\Phi^{1}(t) &= \Phi^{2}(t) = -\tilde{g}\varphi_{00}^{3}\upsilonT(t),\label{Phia_approx} \\			 \tilde{\Phi}^{\alpha\beta}(t) &= -\left( \tilde{g}_{\alpha \alpha} + \tilde{g}_{\alpha \beta} \right)\varphi_{00}^{3}\upsilonT(t),\label{Phiab_approx}
\end{align}
where $\upsilonT(t)$ is defined in Eq.~(\ref{upsilonT}).  In Eq.~(\ref{Phiab_approx}), we used that $\upvarphi^{\alpha}_0(0) = \varphi_{00} + O(\epsilon)$ and kept only the lowest-order terms, considering that the source terms are of quadratic order in $\epsilon$.  From Eqs.~(\ref{Phia_approx}) and (\ref{Phiab_approx}), it follows trivially that
\begin{align}
\Phi^{1} - \tilde{\Phi}^{12} &= -(\tilde{g}-\tilde{g}_{11}-\tilde{g}_{12}) \varphi_{00}^3\upsilonT(t),\label{DeltaPhi1}\\
	\Phi^{2} - \tilde{\Phi}^{21} &= -(\tilde{g}-\tilde{g}_{22}-\tilde{g}_{12}) \varphi_{00}^3\upsilonT(t).\label{DeltaPhi2}
\end{align}

For our particular choice of the hyperfine levels of $^{87}$Rb, the $s$-wave scattering lengths for the processes $|1\rangle|1\rangle \rightarrow |1\rangle|1\rangle$, $|1\rangle|2\rangle \rightarrow |1\rangle|2\rangle$, and $|2\rangle|2\rangle \rightarrow |2\rangle|2\rangle$ are $a_{11}=100.40 a_0$, $a_{12}=97.66 a_0$, and $a_{22}=95.00 a_0$~\cite{mertes07a}, with $a_0$ being the Bohr
radius.  Thus the inter-mode coupling is given approximately by the arithmetic mean of the intra-mode coefficients, i.e., $g_{12}\simeq (g_{11}+g_{22})/2$.  This means that
\begin{align}
	g_{11}-g_{12} \simeq (g_{11}-g_{22})/2\equiv\gamma_1,
\end{align}
from which we get
\begin{align}\label{tilde1}
	\tilde{g}-\tilde{g}_{11}-\tilde{g}_{12} &= \frac{1}{2}N(g_{11}-g_{12})\simeq \frac{\gamma_1}{2}N,\\
	\tilde{g}-\tilde{g}_{22}-\tilde{g}_{12} &= \frac{1}{2}N(2g_{11}-g_{12}-g_{22})+g_{22}-g_{11}\simeq \frac{3\gamma_1}{2}N,
\label{tilde2}
\end{align}
where in the final forms we keep only terms that scale with $N$.

We can now put all these results together with Eqs.~(\ref{I0alpha}) and (\ref{Inalphaapprox}) to write the source terms as
\begin{align}
	\tilde{I}_0^{1}(t) &= -\frac{1}{2}\gamma_1 N \varphi_{00}^3\Big\{\left( 2\tilde{g}_{1 1} |\upvarphi_0^{1}(t)|^{2} + \tilde{g}_{1 2} |\upvarphi_0^{2}(t)|^{2} \right) \upsilonT(t) +  \tilde{g}_{1 1} (\upvarphi_0^{1}(t))^{2} \upsilonT^{*}(t) \nonumber\\
	&\qquad + 3 \tilde{g}_{1 2} \upvarphi_0^{1}(t) \left[ \upvarphi_0^{2}(t)\upsilonT^{*}(t) + \upvarphi_0^{2*}(t)\upsilonT(t) \right]\Big\},\\
	\tilde{I}_0^{2}(t) &= -\frac{3}{2}\gamma_1 N \varphi_{00}^3\Big\{ \left( 2\tilde{g}_{2 2} |\upvarphi_0^{2}(t)|^{2} + \tilde{g}_{12} |\upvarphi_0^{1}(t)|^{2} \right) \upsilonT(t) +  \tilde{g}_{2 2} (\upvarphi_0^{2}(t))^{2} \upsilonT^{*}(t) \nonumber\\
	&\qquad + \frac{1}{3}\tilde{g}_{12} \upvarphi_0^{2}(t) \left[ \upvarphi_0^{1}(t)\upsilonT^{*}(t) + \upvarphi_0^{1*}(t)\upsilonT(t) \right]\Big\},\\
	\tilde{I}_n^{1}(t) &= -\frac{1}{2}\gamma_1 N \varphi_{00}^3\frac{\langle \xi_{0}^{3} | \xi_{n} \rangle}{E_n-E_0} e^{-i (E_n-E_0)t/\hbar},\label{In1}\\
	\tilde{I}_n^{2}(t) &= -\frac{3}{2}\gamma_1 N \varphi_{00}^3\frac{\langle \xi_{0}^{3} | \xi_{n} \rangle}{E_n-E_0} e^{-i (E_n-E_0)t/\hbar}.\label{In2}
\end{align}
Because $\gamma_1 \ll g_{22}, g_{12}, g_{11}$, however, we expect the contribution from these source terms to the two-mode evolution to be negligible.  In other words, since the initial longitudinal function~$\varphi_{n0}$ is not much different from the adiabatic function $\tilde\upvarphi^{\alpha}_n(0)$, defined in Eq.~(\ref{vphinadiabatic}), the evolution of $\upvarphi^{\alpha}_n$ should not differ much from the adiabatically tracking solutions.


\subsection{Numerical simulations} 
\label{sub:numerical_results}

To study the two-mode evolution described by our perturbative model, we numerically integrate Eqs.~(\ref{vphi01}) and (\ref{vphi02}) and compare the results against the exact numerical results obtained from evolving the two-mode BEC with the two-mode, coupled, three-dimensional Gross-Pitaevskii equations~(\ref{twomodeGPE})~\cite{xmds}.  We restrict our numerical analysis to the same scenario of the three-dimensional simulations presented in Ref.~\cite{tacla10}; i.e., we consider the two-mode condensate to be trapped by cigar-shaped potentials of the form
\begin{equation}
\label{potential_dynamics}
    V(\rho,z) = \frac{1}{2}(m\omega_T^2 \rho^2 + k z^q),
\end{equation}
with $q=2$, $4$, and $10$.  These three potentials allow us to explore how the results vary due to the inhomogeneity of the trapping potentials.  Notice that the limit $q\rightarrow \infty$ recovers the case of a homogeneous, hard-walled trap.

To discuss the parameters we use in our simulation, we need first to reprise results from Refs.~\cite{boixo09,tacla10} regarding the length and time scales of the trapping potentials, which are set by properties of the ground state when all the atoms are in the hyperfine state $|1\rangle$.  The discussion here only applies to the cigar-shaped traps for which we do numerical simulations.  We let $\rho_0=\sqrt{\hbar/m\omega_T}$ denote the (bare) ground-state width in the transverse directions; $z_0=(\hbar^2/mk)^{1/(q+2)}$ is an approximation to the bare ground-state width in the longitudinal direction.  The ground-state wave function spreads in the longitudinal direction due the repulsive scattering term; a Thomas-Fermi estimate of the longitudinal width when there are $N$ atoms in the trap, given by Eq.~(3.50) of Ref.~\cite{boixo09}, is
\begin{equation}
z_N=z_0\left(\frac{q+1}{q}\frac{N}{\NL}\right)^{1/(q+1)}\;,
\label{zN}
\end{equation}
where $\NL=(\rho_0/2a_{11})(\rho_0/z_0)$ is a lower critical atom number defined by Eq.~(3.27) of Ref.~\cite{boixo09}.  The upper critical atom number $\NT$, at which the nonlinear scattering energy becomes comparable to the transverse kinetic energy, is defined in Eq.~(3.9) of Ref.~\cite{tacla10}:
\begin{equation}
\NT=\frac{q}{q+1}\!\left(\frac{2q+1}{q}\right)^{(q+1)/q}\!\!\left(\frac{z_0}{\rho_0}\right)^{2(q+1)/q}\NL
=\frac{q}{2(q+1)}\!\left(\frac{2q+1}{q}\right)^{(q+1)/q}\!\frac{z_0}{a_{11}}\!\left(\frac{z_0}{\rho_0}\right)^{2/q}.
\label{defNT}
\end{equation}

In our simulations, we set the transverse frequency to 350 Hz, and for a harmonic longitudinal trap ($q=2$), we set the longitudinal frequency, $\sqrt{k/m}$, to 3.5 Hz.  These frequencies are typical of values accessible to experiment~\cite{gorlitz01}, and they give an upper critical atom number $\NT\simeq14\,000$ atoms for $q=2$.  For $q=4$ and $q=10$, we choose the longitudinal stiffness parameter $k$ so that $\NT$ has this same value, thus giving all the traps the same one-dimensional regime of atom numbers.  For these parameters, we find $\rho_0\simeq 0.6\,\mu$m, and the aspect ratio of the bare traps ($\rho_0$:$z_0$) to be approximately 1:10, 1:24, and 1:57, respectively, for $q=2,4,10$.  When the traps are loaded with $N = \mathcal{N}_T$ atoms, the condensate aspect ratios $(\rho_0$:$z_N)$ become 1:158, 1:146, and 1:138 for $q = 2,4,10$.  These parameters are typical of those in elongated BECs~\cite{jo}.

The role of the upper critical atom number can be appreciated by displaying explicitly the dimensionless expansion parameter for our perturbation theory.  Inspection of Eqs.~(\ref{vphi01}) and (\ref{vphi02}) shows that this parameter is given by the ratio
\begin{align}
	\frac{N^2 g_{11}^2\UpsilonT/z_N^2}{g_{11}\eta_TN/z_N}
\sim\frac{1}{3}\left(\frac{N}{\NT}\right)^{q/(q+1)},
\end{align}
where the numerator is the characteristic size of the $\epsilon^2$ terms multiplying $\varphi_0^1$ and $\varphi_0^2$ in Eqs.~(\ref{vphi01} and (\ref{vphi02}), and the denominator is the characteristic size of the scattering interaction in the $\epsilon$ term in these same equations.  The form on the right gives the scaling with $N$ and a judicious estimate of the constant multiplying this scaling, both obtained using Eqs.~(\ref{zN}) and~(\ref{defNT}) and the explicit expression for $\UpsilonT$ for a harmonic transverse trap given in Eq.~(\ref{Upsilon_harmonic}).  The scaling indicates that the perturbation theory should work better for large $q$, i.e., for hard-walled longitudinal traps.


\subsection{Differential phase dynamics: nonlinear Ramsey interferometry}
\label{Ramsey}

We are now in position to revisit the numerical simulations of a nonlinear Ramsey interferometer presented in Ref.~\cite{tacla10} and use the results of the time-dependent relative-state decomposition to analyze the relative-phase dynamics of the two-mode condensate.

As in typical Ramsey interferometry schemes, the protocol runs as
follows.  The atoms are first condensed to the state
$\psi({\mathbf r})|1\rangle$, and a fast optical pulse, performing a $\pi/2$ rotation about the Bloch $y$ axis, suddenly creates
the superposition state $\psi({\mathbf r})(|1\rangle +
|2\rangle)/\sqrt{2}$ for each atom. The atoms are then allowed to
evolve freely for a time~$t$, which brings the atomic state
to [$\psi^{1}({\mathbf r},t)|1\rangle +
\psi^{2}({\mathbf r},t)|2\rangle]/\sqrt{2}$. A second transition
between the hyperfine levels is then used to transform any differential phases
between the two modes into population information that is finally
detected. For this second transition, we choose a $\pi/2$ rotation about
the Bloch $x$ axis, changing the atomic state to
\begin{equation}
    \label{eq:GP_finalstate}
    \frac{1}{2}\bigl(\psi^{1}-i\psi^{2}\bigr)|1\rangle-
    \frac{i}{2}\bigl(\psi^{1}+i\psi^{2}\bigr)|2\rangle.
\end{equation}
Thus any differential phases accumulated by the mean-field wave functions will give rise to an interference fringe pattern in the detection probabilities for each
hyperfine level,
\begin{equation}
    \label{eq:probs}
    p_{1,2}=\frac{1}{2}\bigl[1\mp
    \mbox{Im}(\langle\psi^{2}|\psi^{1}\rangle)\bigr],
\end{equation}
due to the overlap of the two spatial wave functions,
\begin{equation}
\langle\psi^{2}|\psi^{1}\rangle=
\int d^3 r\,\psi^{2*}({\bf r}, t)\psi^{1}({\bf r}, t).
\label{overlaponetwo}
\end{equation}
Note that the imaginary part of the overlap~(\ref{overlaponetwo}) is responsible for the fringe
pattern in this interferometry scheme.

\subsubsection{Time scales.} 
\label{sub:time_scales}

The relevant time scale for the relative-phase dynamics of the two-mode condensate can easily be estimated if one completely neglects the spatial evolution of the condensate.  Under this approximation, the two-mode evolution becomes trivial.  Due to the difference in scattering lengths, the only effect of the evolution is to introduce a differential phase shift between the states $|1\rangle$ and $|2\rangle$.  This relative phase simply corresponds to the difference between the average scattering energies of the condensate, which implies that the detection probabilities~(\ref{eq:probs}) oscillate as
\begin{equation}
\label{idealsignal}
 p_{1,2}=\frac{1}{2}\bigl[1\mp\sin(\Omega_N t)\bigr],
\end{equation}
where
\begin{equation}
\Omega_N \simeq N\eta\gamma_1/\hbar
\label{OmegaN}
\end{equation}
is the fringe frequency, keeping only terms that scale with $N$.  Here the quantity
\begin{equation}
    \label{eq:eta}
    \eta = \int d^3r\,|\psi({\bf r})|^4
\end{equation}
is a measure of the inverse volume occupied by the ground-state wave
function. This quantity can also be thought of as the average density per atom; i.e., $N\eta$ is the density, $N|\psi({\bf r})|^2$, averaged over the probability density $|\psi({\bf r})|^2$.  This fringe pattern allows one to
estimate the coupling constant $\gamma_1$ with an uncertainty given
by
\begin{equation}
 \delta\gamma_1 =
 \frac{\langle(\Delta\hat{J}_y)^2\rangle^{1/2}}{|d\langle\hat{J}_y\rangle/d\gamma_1|}
 \sim \frac{1}{\sqrt{N}N\eta},\label{deltagamma1}
\end{equation}
whose scaling with the atom number is better than the optimal $1/N$ scaling of linear interferometry~\cite{tacla10}.

Within the standard reduced-dimension approximation of Eq.~(\ref{dDGPE}), the dominant contribution to the fringe frequency is
\begin{equation}
\Omega_{\rm rda}=N\eta_T\eta_L\gamma_1/\hbar,
\label{Omegarda}
\end{equation}
where
\begin{equation}
\eta_L=\int d^{\,d}r\,|\varphi_{00}({\bm r})|^4.
\end{equation}
This frequency sets the characteristic time scale of the short-time relative evolution of the longitudinal wave functions $\upvarphi_0^{1}$ and $\upvarphi_0^{2}$.

\begin{figure}[htbp]
	\centering
		\includegraphics[scale=.5]{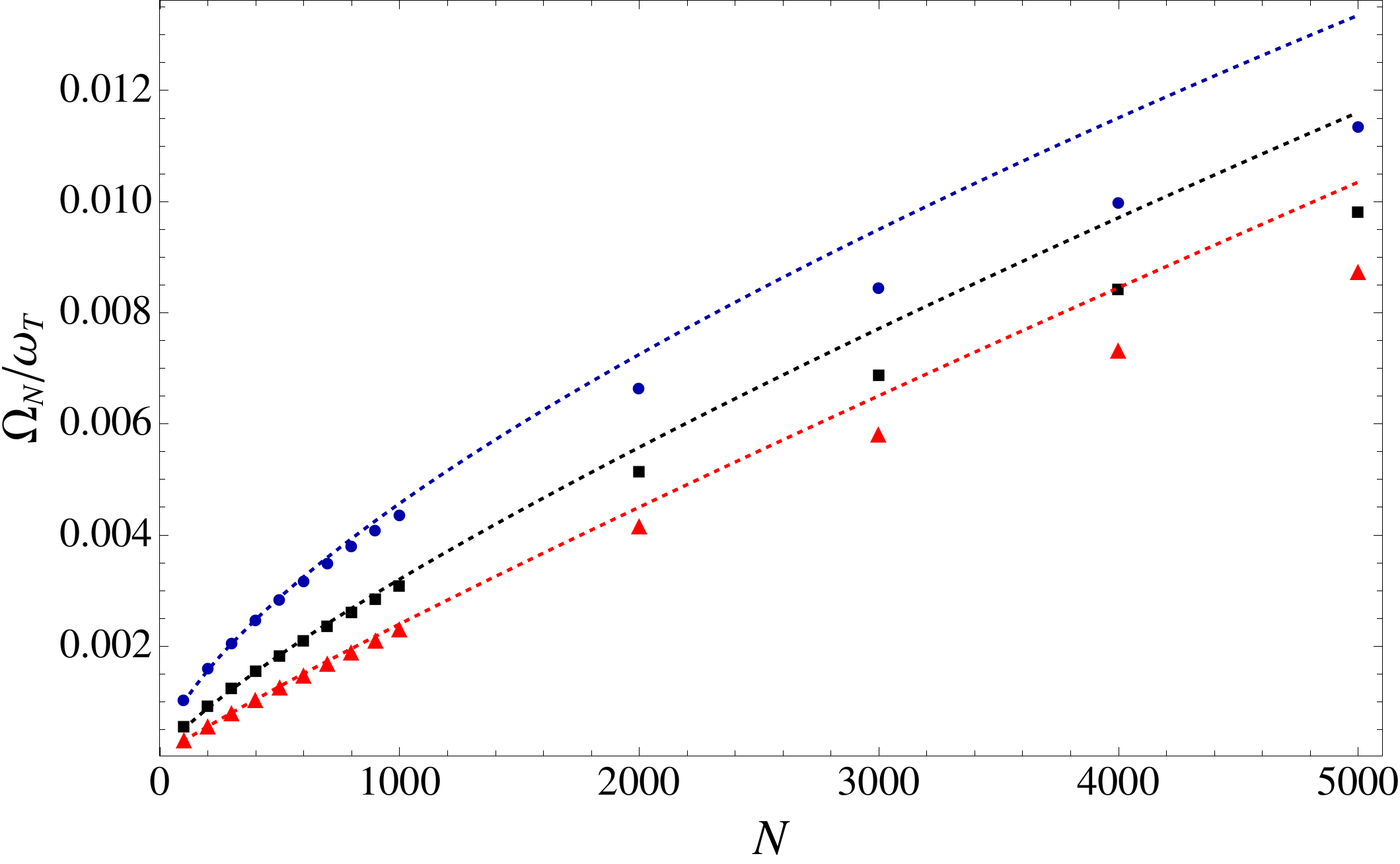}
	\caption{Idealized fringe frequency $\Omega_N=N\eta\gamma_1/\hbar$, in units of the bare transverse frequency $\omega_T$, as a function of the number of condensed atoms.  This fringe frequency, calculated using the numerical results for $\eta$, is plotted as circles (blue) for $q=2$, squares (black) for $q=4$, and triangles (red) for $q=10$.  The dashed lines correspond to the standard reduced-dimension estimate, $\Omega_{\rm rda}=N\eta_T\eta_L\gamma_1/\hbar$, which gives an upper-bound to the fringe frequency $\Omega_N$ calculated using the numerical results for $\eta$.  Both frequencies are much smaller than the transverse trap frequency $\omega_T$.}
	\label{fig:OmegaN}
\end{figure}

In Fig.~\ref{fig:OmegaN}, we plot the values of the Ramsey fringe frequencies $\Omega_N$ and $\Omega_{\rm rda}$, as a function of atom number and the three different values of $q$.  We calculate $\Omega_N$ using the numerical evaluation of $\eta$.  Not surprisingly, $\Omega_{\rm rda}$ gives an upper bound on the numerical values of $\Omega_N$.  The important point here is not the difference between $\Omega_N$ and $\Omega_{\rm rda}$, but rather that both are much smaller than the transverse frequency $\omega_T$, as is to be expected for the highly anisotropic potentials used in our simulations.  This result indicates that the longitudinal frequencies do satisfy the adiabatic condition $\Omega_N,\Omega_{\rm rda} \ll \omega_T$.  We therefore expect that the longitudinal two-mode equations~(\ref{vphi01}) and (\ref{vphi02}) can be used to model the dynamics of the BEC interferometry protocol.

\begin{figure}[!t]
	\centering
		\includegraphics[width=\textwidth]{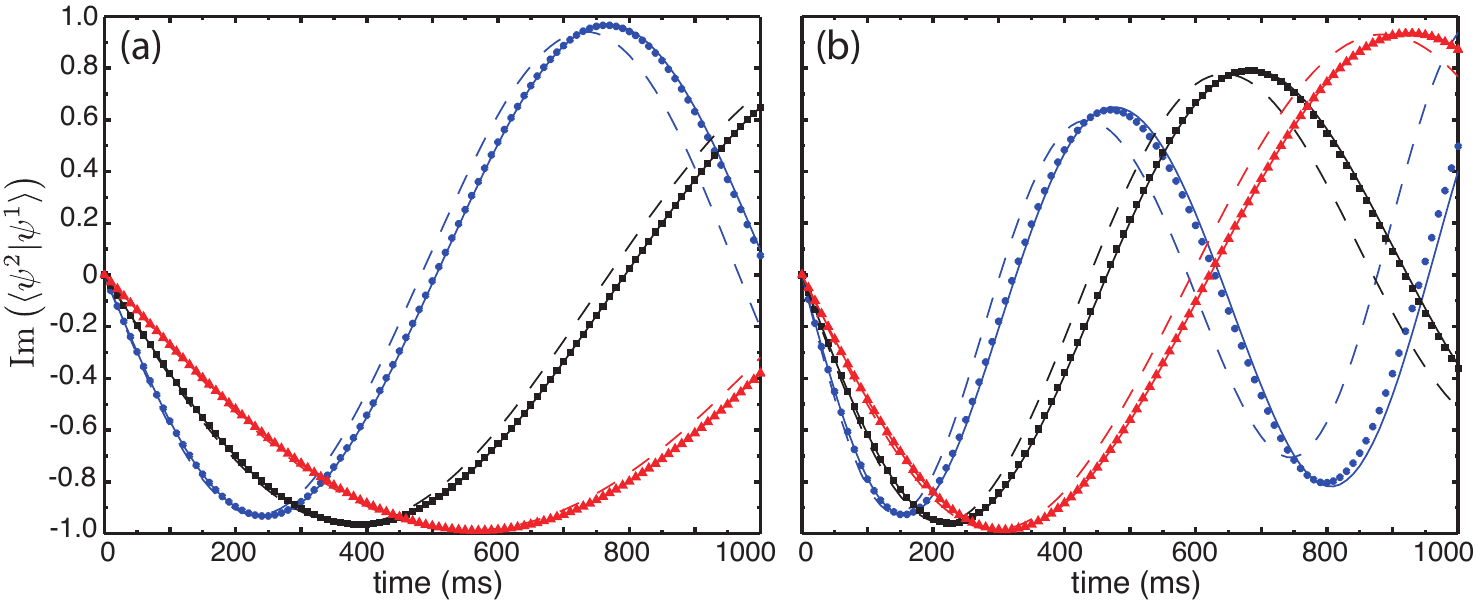}
	\caption{Ramsey fringes for a cigar-shaped $^{87\!}{\rm Rb}$ BEC of (a)~500 and (b)~1\,000~atoms. The points correspond to the results of the numerical integration of the time-dependent, coupled, two-mode, three-dimensional GP equations~(\ref{twomodeGPE}) for the different trapping potentials~(\ref{potential_dynamics}): circles (blue) correspond to $q = 2$, squares (black) to $q = 4$, and triangles (red) to $q = 10$.  The corresponding solid lines are the numerical results of the integration of Eqs.~(\ref{vphi01}) and (\ref{vphi02}), given by the relative-state-decomposition perturbation theory.  The dashed lines represent the results coming from the numerical integration of the time-dependent, two-mode, quasi-1D GP equation (the standard reduced-dimension approximation).  For all three trap geometries, the agreement between our perturbation theory and the exact 3D numerics is remarkably good during the whole 1s of integration time.  The reduced-dimension approximation, on the other hand, can only account for earlier stages of the evolution.  As $q$ increases, both models achieve a better performance.  Note that the reduction in the fringe visibility is small, being almost absent for $q=10$.  Both fringe frequency and amplitude are better predicted by our model than by the standard reduced-dimension (quasi-1D) approximation.}
	\label{fig:OverlapQuintic-500-1000}
\end{figure}


\subsubsection{Ramsey fringes.} 
\label{subsub:ramsey_fringes}

According to the relative-state decomposition~(\ref{timedependentpsiRSD}), it is easy to see that, at the order we are working, the dominant longitudinal functions $\upvarphi^{1}_0$ and $\upvarphi^{2}_0$ carry all the information about the relative phase between the three-dimensional condensate wave functions $\psi^{2}$ and $\psi^{1}$, since
\begin{equation}
\langle\psi^{2}|\psi^{1}\rangle
=\langle\upvarphi^{2}_0|\upvarphi^{1}_0\rangle + O(\epsilon^2).\label{overlapRSD}
\end{equation}
By using the numerical solutions of the time-dependent, coupled, 3D GP equations, we calculate
the spatial overlap $\langle\psi^{2}|\psi^{1}\rangle$ and by using the numerical solutions of Eqs.~(\ref{vphi01}) and (\ref{vphi02}), we calculate the overlap $\langle\upvarphi^{2}_0|\upvarphi^{1}_0\rangle$, both as a function of time for various atom numbers.  In addition, we calculate the spatial overlap between the numerical solutions of the quasi-1D, time-dependent GP equation (the standard reduced-dimension approximation) to use as a benchmark for our relative-state-decomposition perturbation model~\cite{note}. We stress that these spatial overlaps and the resulting Ramsey fringes, including those coming from the standard reduced-dimension approximation, are more than the simple estimates~(\ref{OmegaN}) and~(\ref{Omegarda}), since the numerical integrations include effects of position-dependent phase shifts and reduction in fringe visibility (see Sec.~\ref{subsub:thomas_fermi_estimates}) not captured by the simple estimates of fringe frequency.

\begin{figure}[htb!]
	\centering
		\includegraphics[width=\textwidth]{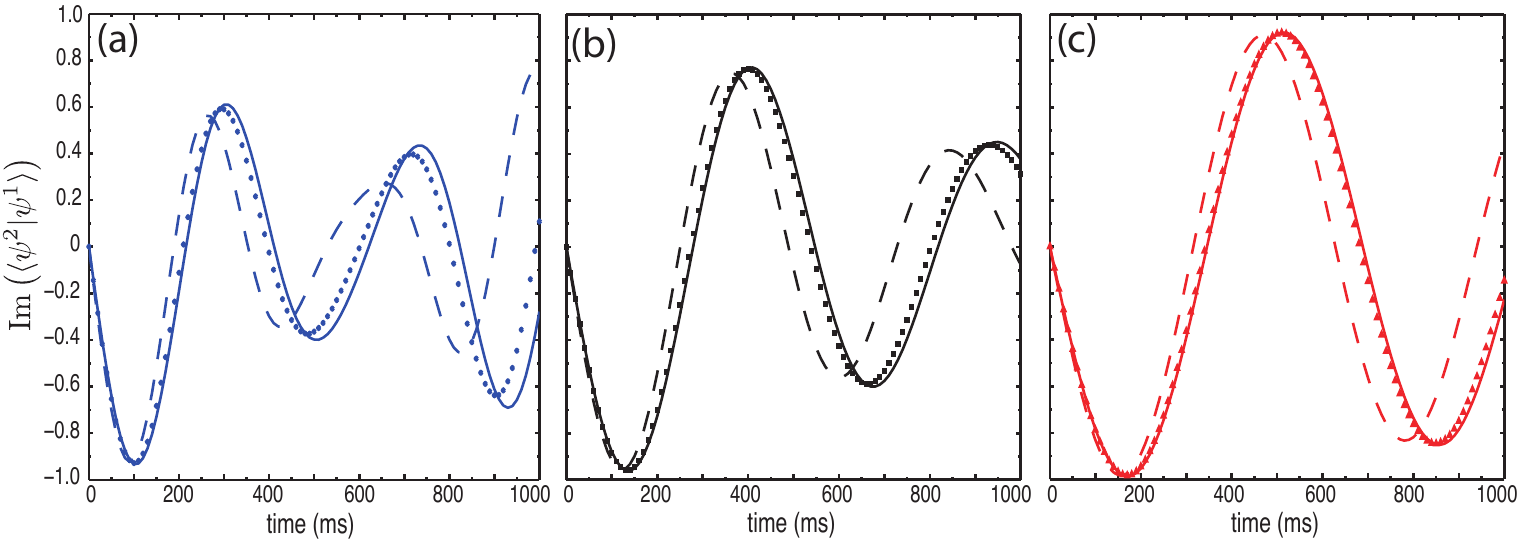}
	\caption{Ramsey fringes for a cigar-shaped $^{87\!}{\rm Rb}$ BEC of 2\,000 atoms trapped by the potentials~(\ref{potential_dynamics}): (a)~$q = 2$, (b)~$q = 4$, (c)~$q = 10$.  The points represent the numerical results of the integration of the time-dependent, coupled, two-mode 3D GP equations.  Solid lines are the corresponding relative-state predictions, whereas the dashed lines are results coming from the numerical solution of the quasi-1D GP equation.  The reduction in fringe visibility is still well captured by our model, with agreement with the 3D numerics getting better as $q$ increases. The predicted fringe frequency, however, is a bit too small, which points to the breakdown of the perturbation theory.}
	\label{fig:overlap2000}
\end{figure}

In Fig.~\ref{fig:OverlapQuintic-500-1000}(a), we compare the Ramsey fringes obtained from the imaginary part of these three spatial overlaps for a condensate of 500 atoms.  For all three longitudinal potentials, our perturbation theory reproduces the exact 3D numerics remarkably well during the entire integration time interval.  The standard reduced-dimension approximation, on the other hand, only holds in the earliest stages of the evolution, for about a quarter of a fringe period; note that the deviations are quite significant in spite of the relatively small number of atoms in the condensate ($N/\NT\sim 3.5 \times 10^{-2}$).  It is no surprise that as the traps get harder and more homogeneous, both models achieve a better performance.  In this case of rather small nonlinearities, the reduction in the fringe visibility due to position-dependent differential phases is quite small, being almost absent for $q=10$.  This effect becomes obvious for stronger couplings, however, as illustrated in Fig.~\ref{fig:OverlapQuintic-500-1000}(b) for a condensate of $1\,000$ atoms.  For this many atoms and the same time window of $1\,$s, the agreement between our perturbation model and the 3D solution is still quite good, especially for the more homogeneous trapping potentials $q=4$ and $q=10$.  Both fringe frequency and amplitude are well predicted by our model, showing a substantial improvement over the standard reduced-dimension approximation.  For a harmonic trap, small deviations from the expected signal accumulate over time, resulting in an evident phase mismatch around $t=500$ ms.

For a condensate of 2\,000 atoms, our model starts to break down, as shown in Fig.~\ref{fig:overlap2000}.  For all $q$'s, the reduction in fringe visibility is still well captured by our perturbative equations, with agreement with the 3D numerics getting better as $q$ increases, but the predicted fringe frequency is a bit too small.

\begin{figure}[htbp]
	\centering
		\includegraphics[width=\textwidth]{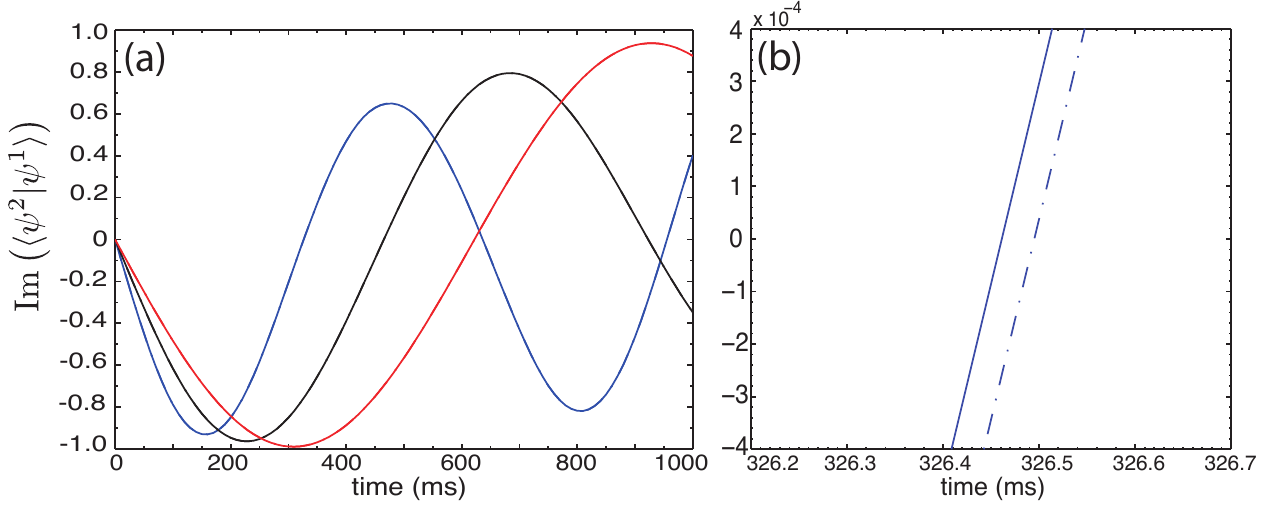}
	\caption{(a) Ramsey fringes for 1\,000 atoms using the solutions of Eqs.~(\ref{vphi01}) and (\ref{vphi02}) with (solid line) and without (dash-dotted line) the contribution from the source terms for $q = 2, 4, 10$ (blue, black, red).  The two lines lie almost on top of one another. (b)~Closeup for the $q=2$ case around $t=326.5$ ms, which shows a difference of the order of $10^{-4}$ between the two fringes.}
	\label{fig:OverlapQuinticNoSource1000}
\end{figure}


\subsubsection{Source terms.} 
\label{sub:source_terms}

As a brief remark, we point out that in all the results presented in this section, the solutions to Eqs.~(\ref{vphi01}) and (\ref{vphi02}) take into account the source terms that arise due to the difference $\upvarphi^{\alpha}_n(0)-\tilde{\upvarphi}^{\alpha}_n(0)$.  We do expect that the contribution of the source terms is quite small for the initial condition we use.  To verify this, we plot in Fig.~\ref{fig:OverlapQuinticNoSource1000}(a) the predicted Ramsey fringes for 1\,000 atoms using the solutions of Eqs.~(\ref{vphi01}) and (\ref{vphi02}) with and without the contribution from the source terms.  The difference in the computed overlaps is indeed quite small, being of order $10^{-4}$ for the case of 1\,000 atoms  and $q=2$, as shown in  Fig.~\ref{fig:OverlapQuinticNoSource1000}(b).


\subsubsection{Thomas-Fermi estimates.} 
\label{subsub:thomas_fermi_estimates}

For the traps and atom numbers that we are considering, it is legitimate
to ignore the kinetic-energy term in
Eqs.~(\ref{vphi01}) and (\ref{vphi02}) and work in the longitudinal Thomas-Fermi approximation.  Within this regime, the probability densities do not change with time, i.e.,
\begin{equation}
|\upvarphi_0^{\alpha}(z,t)|^2=|\upvarphi_0(z,0)|^2,
\end{equation}
with $|\upvarphi_0(z,0)|^2$ given by the Thomas-Fermi solution of Eq.~(\ref{quinticequation}),
\begin{equation}
\label{newTF}
	|\upvarphi_0(z,0)|^2 = \frac{\tilde{\mu}_L-V_L}{\tilde{g}\eta_T} + \epsilon \frac{3\tilde{g}\UpsilonT}{\eta_T} \left( \frac{\tilde{\mu}_L-V_L}{\tilde{g}\eta_T} \right)^2
\equiv \mathcal{Q}_0(z).
\end{equation}

\begin{figure}[htbp]
	\centering
		\includegraphics[width=\textwidth]{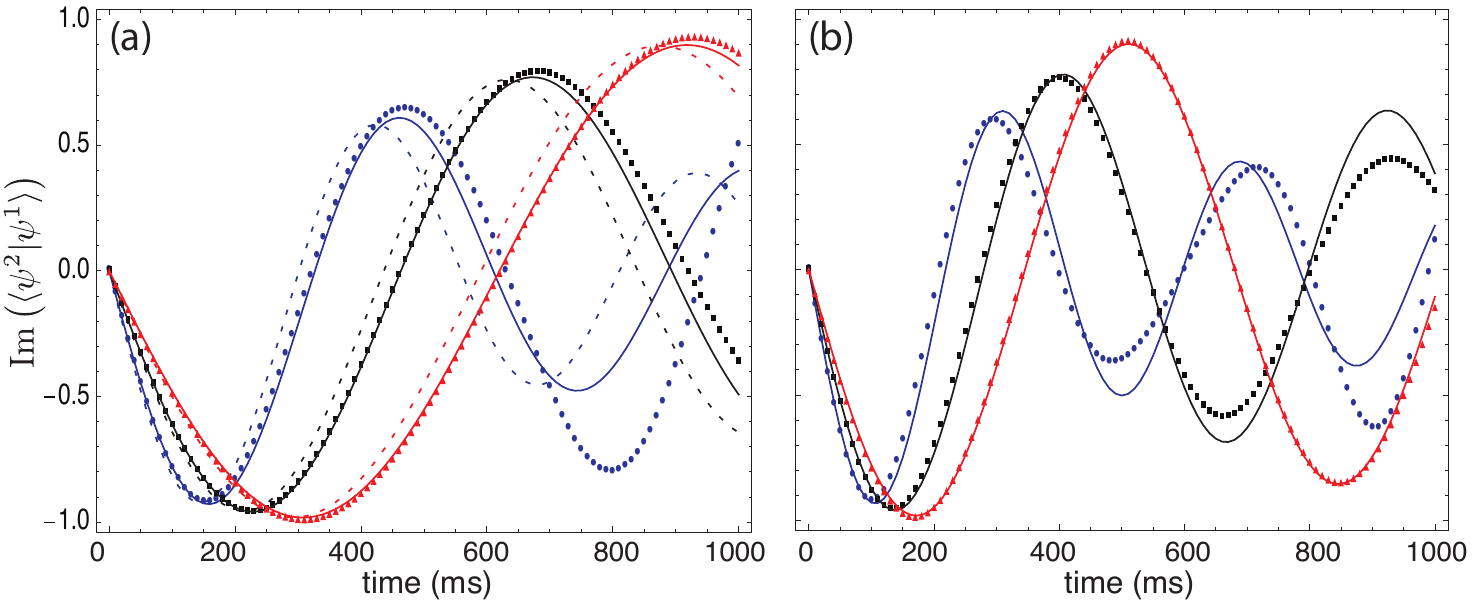}
	\caption{(a)~Ramsey fringes, as in
        Figs.~\ref{fig:OverlapQuintic-500-1000}(b) and~\ref{fig:overlap2000}, for $q = 2, 4, 10$ (blue, black, red), but here the solid lines correspond to the Thomas-Fermi estimate~(\ref{overlaplongTF}).  (a)~1\,000 atoms.  The Thomas-Fermi estimate provides a good account of the exact 3D fringe signal (circles, squares, triangles) for short times and harder traps, far superior to the predictions (dashed lines) of the quasi-1D model.  For longer times, however, the analytical Thomas-Fermi model breaks down, reflecting the fact that the probability densities can no longer be regarded as constants in time.
        (b)~2\,000 atoms, with the ineffective quasi-1D results omitted for clarity.  The agreement between the exact 3D overlap and the Thomas-Fermi estimate is still reasonably good for short times, but deviations from the exact fringe pattern become significant for $q=2$ after a quarter of a fringe period and for $q=4$ after a full period, due to changes in the probability densities.  For $q=10$, however, the model provides a correct prediction for the entire integration time.}
	\label{fig:Signal_Quintic_TF-1000-2000}
\end{figure}

Under this approximation the two-mode evolution can be described by a simple accumulation of a phase that depends on the local atomic linear density, given by
\begin{align}
    \label{eq:upvarphiTF}
    \upvarphi_0^{\alpha}(z, t) =
    \sqrt{\mathcal{Q}_0(z)}\exp\!\bigg[-\frac{it}{\hbar}\bigg(&\frac{1}{2} k z^q +
    \eta_T \mathcal{Q}_0(z)(\tilde{g}_{\alpha \alpha} + \tilde{g}_{\alpha \beta})\nonumber\\
	&- \epsilon\UpsilonT \mathcal{Q}^2_0(z)(3 \tilde{g}_{\alpha \alpha}^2 + 3 \tilde{g}_{\alpha \beta}^2 + \tilde{g}_{\alpha \beta} (4\tilde{g}_{\alpha \alpha} + 2\tilde{g}_{\beta \beta})\bigg) \bigg].
\end{align}
This yields an overlap
\begin{equation}
\label{overlaplongTF}
\langle\upvarphi_0^{2}|\upvarphi_0^{1}\rangle
=\int dz\,\mathcal{Q}_0(z) e^{-i\updelta\uptheta(z)}.
\end{equation}
The position-dependent differential phase shift is
\begin{equation}
\label{updeltatheta}
\updelta\uptheta(z)=\frac{N\eta_T \mathcal{Q}_0(z) \gamma_1t}{\hbar} \bigg(1 -\epsilon \frac{\UpsilonT}{\eta_T} \mathcal{Q}_0(z) (3 \tilde{g}_{11} + 3\tilde{g}_{22} + 2 \tilde{g}_{12})\bigg),
\end{equation}
where in the leading-order term, we keep only the terms that scale with $N$, as in Eqs.~(\ref{tilde1}) and (\ref{tilde2}).

In Fig.~\ref{fig:Signal_Quintic_TF-1000-2000}, we show Ramsey fringes coming from the imaginary part of the overlap~(\ref{overlaplongTF}), comparing these Thomas-Fermi fringes with the exact fringes from Figs.~\ref{fig:OverlapQuintic-500-1000} and~\ref{fig:overlap2000}.

As an aside, note that we can easily retrieve the Thomas-Fermi prediction of the quasi-1D approximation by taking the limit $\epsilon\rightarrow 0$.  This gives the following approximation for the overlap~(\ref{overlaponetwo})
\begin{equation}
\label{analyticaloverlap}
	\langle\psi^{2}|\psi^{1}\rangle
\simeq \langle\phi^{2}|\phi^{1}\rangle =\int dz\,
q_0(z) e^{-i\delta\theta(z)},
\end{equation}
where the relative phase is
\begin{equation}
\label{eq:deltatheta}
\delta\theta(z)=\frac{N\eta_T q_0(z) \gamma_1t}{\hbar},
\end{equation}
with $q_0(z)$ being given by
\begin{equation}
\label{thomasfermi1d}
	q_0(z) = \frac{\mu_L-k z^{q}/2}{\tilde{g}\eta_T}.
\end{equation}
Here $\mu_L$, the zero-order longitudinal part of the chemical potential, is determined from the normalization condition for $q_0(z)$, which gives
\begin{equation}
    \mu_L = \frac{k}{2}\left(\frac{q+1}{q}\frac{\tilde g\eta_T}{k}\right)^{q/(q+1)}.
\end{equation}

As far as the spatial overlap~(\ref{overlaplongTF}) is concerned, both the transverse and the longitudinal corrections play a role in rectifying the standard reduced-dimension approximation~(\ref{analyticaloverlap}).  Comparing the two overlaps, one can interpret the longitudinal correction as coming from the distribution $\mathcal{Q}_0(z)$, whereas the transverse correction comes from a rectified $\eta_T$, which is given by
\begin{align}
\label{upetaT}
\upeta_T(z)=\eta_T -\epsilon \UpsilonT \mathcal{Q}_0(z) (3 \tilde{g}_{11} + 3\tilde{g}_{22} + 2 \tilde{g}_{12}).
\end{align}
Note that $\upeta_T(z)$ is a function of the local axial distribution $\mathcal{Q}_0(z)$ and, hence, takes into account inhomogeneities in the trapping confinement.


\subsection{Average densities per atom and density overlap} 
\label{subsub:density_evolution}

\begin{figure}[htbp]
	\centering
		\includegraphics[height=6.7in]{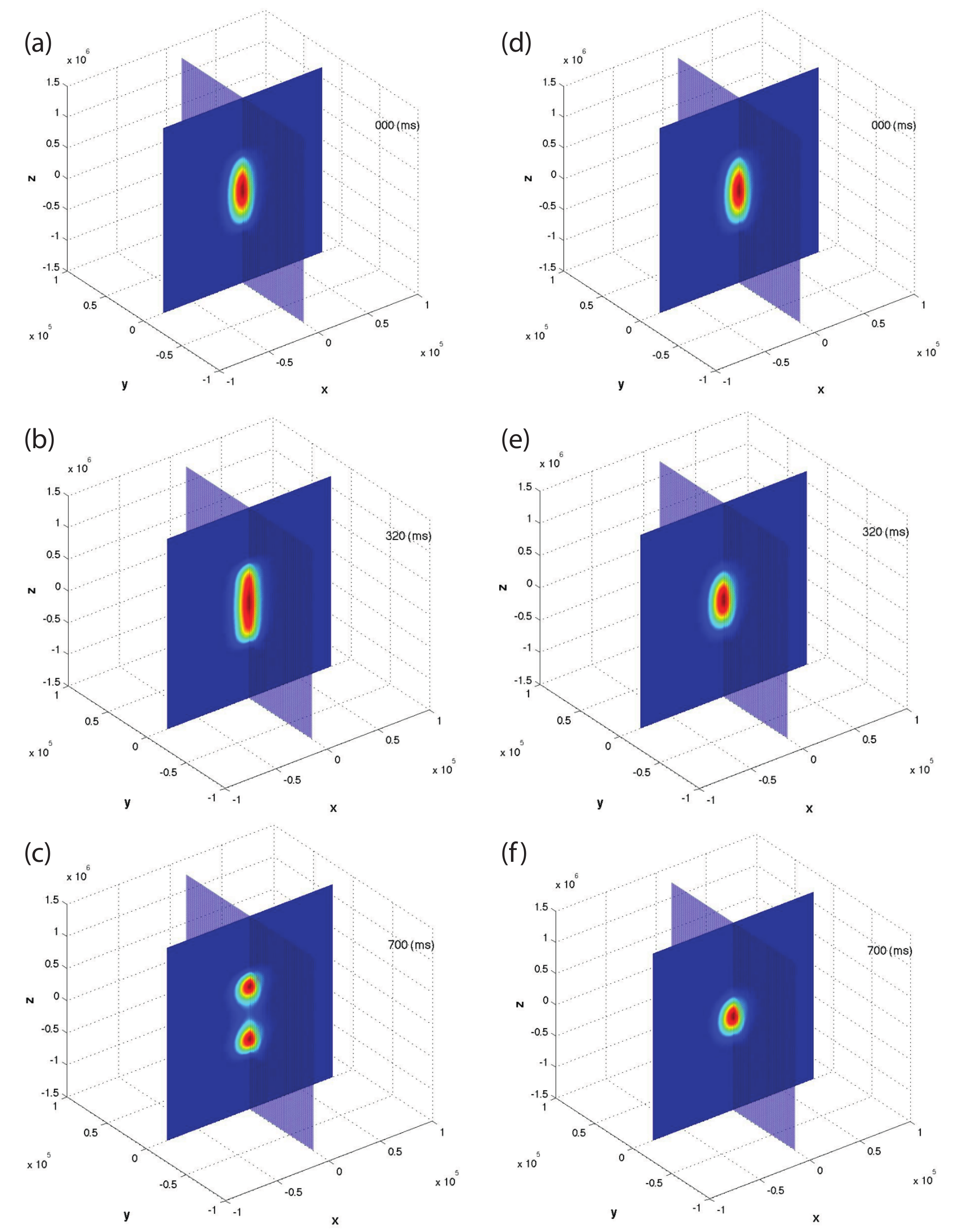}
\caption{Density evolution for a BEC of $1\,000$ atoms and $q=2$. The plots on the left show the time evolution of $|\psi^{1}(\rho,z,t)|^2$, whereas the time evolution of $|\psi^{2}(\rho,z,t)|^2$ is illustrated by the plots on the right.  Spatial co\"ordinates are displayed in 
atomic units.  Initially, both modes have the same density profile as seen in plots~(a) and~(d).  The position-dependent phases acquired by each mode drive differences between the atomic densities associated with the two hyperfine levels, as seen in plots~(b) and~(e) for 320~ms of evolution. Atoms in mode 1 are pushed towards the edges of the trap due to the stronger intra-mode repulsion ($g_{11}>g_{12}>g_{22}$), while atoms in mode 2 get compressed in the center of trap by the atoms in mode 1 because of the inter-mode repulsion $(g_{12}>g_{22})$.  This leads to spatial separation of the two modes, as shown in plots~(c) and~(f) at 700~ms of the evolution. Due to the trap confinement, this dynamics is eventually reversed.  The dynamical relative-state perturbation theory allows us to investigate the density evolution long enough to capture these effects.}
	\label{fig:Densities}
\end{figure}

To examine further the extent to which our perturbative approach correctly describes the two-mode evolution, we study below the dynamics of the average mean-field densities per atom for various atom numbers.  In particular, we can use the perturbation theory to describe the density evolution up to a point where the spatial separation becomes clearly apparent (see Fig.~\ref{fig:Densities}).

According to the perturbative expansion~(\ref{timedependentpsiRSD}), the average single-mode density, normalized to be a density per atom, to first order in $\epsilon$, is given by
\begin{align}
	\eta_{\alpha}(t) & = \int d^3r\, |\psi^{\alpha}(t)|^4 \label{eta_alpha_def}\\
		&=\eta_T \eta_{L}^{\alpha}(t) + 2 \epsilon \sum_{n=1}^\infty \langle \xi_n| \xi_0^{3}\rangle \int dz\, |\upvarphi^{\alpha}_0(t)|^2 \big(\upvarphi^{\alpha*}_0(t) \upvarphi^{\alpha}_n(t) + \mbox{c.c.}\big),
\label{eta_alpha}
\end{align}
where
\begin{equation}
	\eta_{L}^{\alpha}(t) = \int dz\, |\upvarphi^{\alpha}_0(t)|^4.
\end{equation}
Within the adiabatic approximation given by Eqs.~(\ref{vphin1}) and (\ref{vphin2}), we can split Eq.~(\ref{eta_alpha}) into two terms, a dominant term and a second term that comes from the source term in those equations,
\begin{equation}
	\eta_{\alpha}(t) = \tilde{\eta}_{\alpha}(t) + \tilde{\eta}_{I}^{\alpha}(t).\label{eta_twoterms}
\end{equation}

The dominant term, given by
\begin{equation}
	\tilde{\eta}_{\alpha}(t) = \eta_T \eta_{L}^{\alpha}(t) + 2 \epsilon \sum_{n=1}^\infty \langle \xi_n| \xi_0^{3}\rangle \int dz\, |\upvarphi^{\alpha}_0(t)|^2 \big(\upvarphi^{\alpha*}_0(t) \tilde{\upvarphi}^{\alpha}_n(t) + \mbox{c.c.}\big),
\end{equation}
includes the contributions that arise from $\upvarphi^{\alpha}_0$ and from $\tilde{\upvarphi}^{\alpha}_n$, which is defined by Eq.~(\ref{vphinadiabatic}).  Introducing the quantities
\begin{align}
	\eta_{L6}^{\alpha}(t)&=\int dz\, |\upvarphi^{\alpha}_{0}(t)|^{6},\\
	\eta_{L}^{\alpha \alpha \beta}(t)&=\int dz\, |\upvarphi^{\alpha}_{0}(t)|^{4}|\upvarphi^{\beta}_{0}(t)|^{2},
\end{align}
we can write the dominant term in the more compact form
\begin{align}
	\tilde{\eta}_{\alpha}(t) = \eta_T \eta_{L}^{\alpha}(t) - 4 \epsilon \UpsilonT \left( \tilde{g}_{\alpha \alpha}\eta_{L6}^{\alpha}(t)  + \tilde{g}_{\alpha \beta} \eta_{L}^{\alpha \alpha \beta}(t) \right).
\end{align}

The second term in Eq.~(\ref{eta_twoterms}),
\begin{equation}
	\tilde{\eta}_{I}^{\alpha}(t) = 2 \epsilon \sum_{n=1}^\infty \langle \xi_n| \xi_0^{3}\rangle \int dz\, |\upvarphi^{\alpha}_0(t)|^2 \big(\upvarphi^{\alpha*}_0(t) \tilde{I}^{\alpha}_n(t) + \mbox{c.c.}\big),
\end{equation}
comes from the source term~$\tilde{I}_n^{\alpha}$.  Together with Eqs.~(\ref{In1}) and~(\ref{In2}), this gives
\begin{align}
	\tilde{\eta}_{I}^{1}(t) &= - \epsilon\gamma_1 N \int dz\,\varphi_{00}^3 |\upvarphi^{1}_0(t)|^2 \Big(\upvarphi^{1*}_0(t)\upsilonT(t) + \mbox{c.c.}\Big), \\
	\tilde{\eta}_{I}^{2}(t) &= -3 \epsilon\gamma_1 N \int dz\,\varphi_{00}^3 |\upvarphi^{2}_0(t)|^2 \Big(\upvarphi^{2*}_0(t)\upsilonT(t) + \mbox{c.c.}\Big).
\end{align}

A similar set of steps, using Eqs.~(\ref{timedependentpsiRSD}) and (\ref{vphinapprox}), splits the density overlap, normalized in the same way as the average densities,
\begin{align}
	\eta_{12}(t) & = \int d^3r\, |\psi^{1}(t)|^2|\psi^{2}(t)|^2, \label{eta_12_def}
\end{align}
into two terms,
\begin{align}
	\eta_{12}(t) &= \tilde{\eta}_{12}(t) + \tilde{\eta}_{I}^{12}(t),\label{eta12_twoterms}
\end{align}
where
\begin{align}
	\tilde{\eta}_{12}(t) = \eta_T\eta_L^{12}(t)-2 \epsilon \UpsilonT [(\tilde{g}_{11}+\tilde{g}_{12})\eta_L^{112}(t) + (\tilde{g}_{22}+\tilde{g}_{12})\eta_L^{221}(t)],
\end{align}
\begin{equation}
	\eta_{L}^{12}(t) = \int dz\, |\upvarphi^{1}_0(t)|^2|\upvarphi^{2}_0(t)|^2,
\end{equation}
and
\begin{align}
	\tilde{\eta}_{I}^{12}(t) = -\frac{3}{2}& \epsilon\gamma_1 N \int dz\,\varphi_{00}^3 |\upvarphi^{1}_0(t)|^2 \Big(\upvarphi^{2*}_0(t)\upsilonT(t) + \mbox{c.c.}\Big) \nonumber\\
	& - \frac{1}{2} \epsilon\gamma_1 N \int dz\,\varphi_{00}^3 |\upvarphi^{2}_0(t)|^2 \Big(\upvarphi^{1*}_0(t)\upsilonT(t) + \mbox{c.c.}\Big).
\end{align}

\begin{figure}[htb!]
	\centering
		\includegraphics[width=15.6cm]{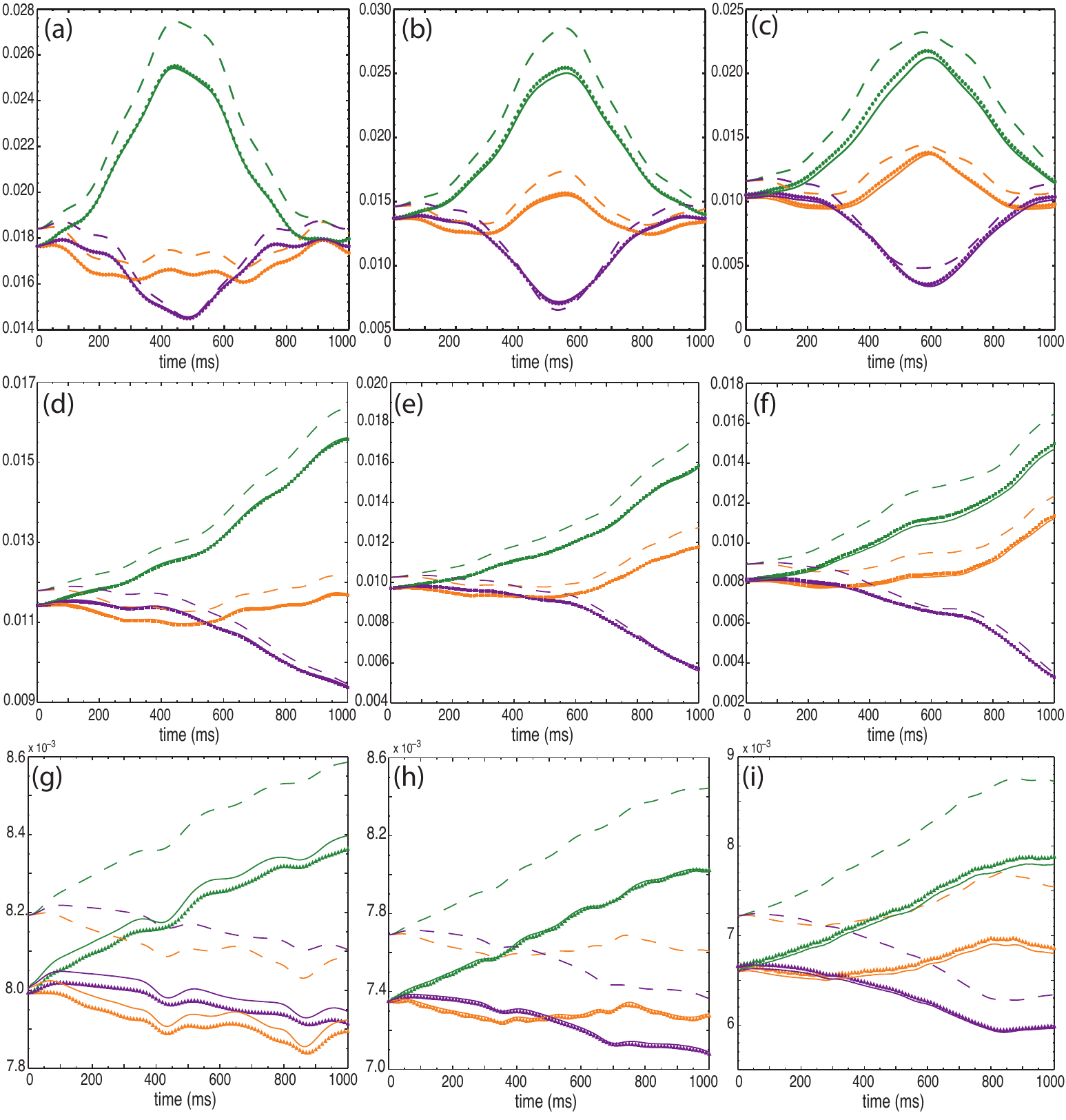}
	\caption{Time evolution of the average densities per atom, $\eta_1$ (orange) and $\eta_2$ (green), and of the density overlap, $\eta_{12}$ (purple), for $q=2$ (top row), 4 (middle row), 10 (bottom row) and $N= 500$~(left column), 1\,000 (middle column), 2\,000 (right column). The points represent the numerical results of the integration of the time-dependent, coupled, two-mode 3D GP equations. Solid lines are the respective relative-state predictions, whereas the dashed lines are results coming from the numerical solution of the quasi-1D GP equation (standard reduced-dimension approximation).  In all cases, the 3D average densities are poorly described by the reduced-dimension approximation.  Our perturbative model, on the other hand, performs extremely well. Note that for 2\,000 atoms, small deviations appear. The inaccurate density dynamics seen in (g) is due to numerical instabilities that affected the calculation of the initial longitudinal wave functions $\upvarphi^1_0(0)$ and $\upvarphi^2_0(0)$.  This numerical error, however, does not seem to affect the good agreement for the relative phases shown in Fig.~\ref{fig:OverlapQuintic-500-1000}(a).}
	\label{fig:Etas_t}
\end{figure}

Our first step in studying the time evolution of the mean-field densities is to calculate $\eta_1(t)$, $\eta_2(t)$, and $\eta_{12}(t)$, as instructed by Eqs.~(\ref{eta_alpha_def}) and~(\ref{eta_12_def}), using the numerical solutions of the time-dependent 3D GP equation~(\ref{twomodeGPE}).  We compare these to the relative-state predictions~(\ref{eta_twoterms}) and~(\ref{eta12_twoterms}).  Again, we also calculate the corresponding predictions of the standard reduced-dimension approximation to use as a benchmark to our model.  Figure~\ref{fig:Etas_t} shows these three quantities in harmonic trap units~\cite{units} for $q=2, 4$, $10$ and $N=500$~(a,d,g), $1\,000$~(b,e,h), and $2\,000$~(c,f,i) atoms.  We note that the 3D numerical solutions, the relative-state model, and the standard reduced-dimension approximation all use different initial conditions.

In all cases, we see that the 3D mean densities are poorly described by the standard reduced-dimension approximation.  Our perturbative model, however, manages to reproduce the exact mean densities remarkably well for 500 and $1\,000$ atoms.  Small deviations start to appear for $2\,000$ atoms.  The case $q=10$ and $N=500$ is an exception to the previous statement.  For this particular case, numerical instabilities affected the calculation of the initial longitudinal wave functions $\upvarphi^1_0(0)$ and $\upvarphi^2_0(0)$, which consequently resulted in the inaccurate density dynamics seen in Fig.~\ref{fig:Etas_t}(g).  Note, however, that this numerical error does not seem to affect the good agreement for the relative phases shown in Fig.~\ref{fig:OverlapQuintic-500-1000}(a).

The two-mode density evolution shown in Fig.~\ref{fig:Etas_t} can roughly be described as follows.  Initially all densities are the same, i.e., $\eta_{1}(0)=\eta_{2}(0)=\eta_{12}(0)$; hence, the earlier stages of the dynamics are essentially dictated by the different scattering couplings.  Atoms in mode 1 are initially pushed towards the edges of the trap due to the stronger intra-mode repulsion ($g_{11}>g_{12}>g_{22}$), while atoms in mode 2 get compressed in the center of trap by the atoms in mode 1 because of the inter-mode repulsion $(g_{12}>g_{22})$.  Therefore, the wave function for mode 1 gets wider than the wave function of the second mode.  In other words, the average atomic density per atom $\eta_{1}$ decreases while $\eta_{2}$ increases, and as the two modes are driven apart, the density overlap $\eta_{12}$ drops. Of course, the atomic repulsion is counterbalanced by the trapping potential, so at some point this dynamics is reversed.

\begin{figure}[h!]
	\centering
		\includegraphics[width=\textwidth]{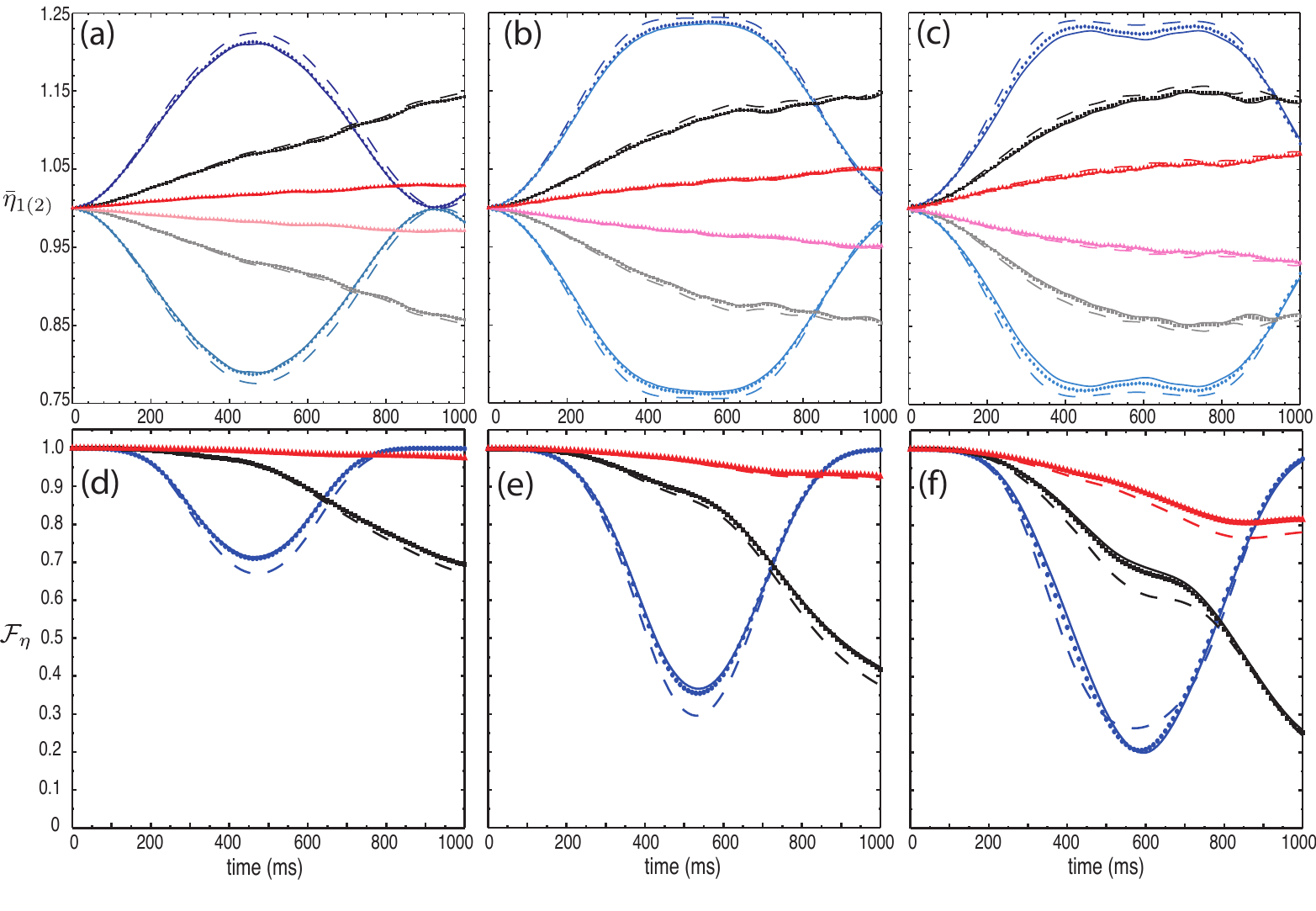}
	\caption{Time evolution of the renormalized average densities per atom, $\bar{\eta}_1$ and $\bar{\eta}_2$ (top row), and the density overlap $\mathcal{F}_{\eta}=\eta_{12}/\sqrt{\eta_{1}\eta_{2}}$ (bottom row) for $q=2,4,10$ (blue, black, red) and $N= 500$ (left column), 1\,000 (middle column), and 2\,000 (right column). For the average densities in the top row, we use a darker color for $\bar{\eta}_1$ and a lighter color for $\bar{\eta}_2$; the results for $\bar{\eta}_1$ lie in the top half of the graph, and for $\bar{\eta}_2$ in the bottom half of the graph. The points represent the numerical results of the integration of the time-dependent, coupled, two-mode 3D GP equations.  Solid lines are the predictions of the relative-state perturbation theory, whereas the dashed lines are results coming from the numerical solution of the quasi-1D GP equation. The agreement between our model and the 3D numerics is very good for the entire integration time. Both the renormalized average densities per atom and the density overlap clearly show the spatial separation of the two modes, which is suppressed as the traps become more homogeneous.}
	\label{fig:Eta1_Eta2}
\end{figure}

The complex nonlinear evolution in Fig.~\ref{fig:Etas_t} makes it challenging to analyze further the dynamical behavior of the mean-field densities.  To gain a deeper understanding of the results, we examine the renormalized single-mode mean densities, $\bar{\eta}_{1}$ and $\bar{\eta}_{2}$, which we define according to
\begin{equation}
	\bar{\eta}_{\alpha} = \frac{\eta_{\alpha}}{\bar{\eta}},
\end{equation}
where
\begin{equation}
	\bar{\eta}=\frac{\eta_1+\eta_2}{2}
\end{equation}
is the arithmetic mean of the two average densities per atom.

Figures~\ref{fig:Eta1_Eta2}(a--c) show the time evolution of $\bar{\eta}_{1}$ and $\bar{\eta}_{2}$ for $q=2, 4$, and $10$.  In each figure, we plot the cases $N=$ 500, 1\,000, and 2\,000.  The time evolution of $\bar{\eta}_{1}$ and $\bar{\eta}_{2}$ is, in fact, much simpler than that of $\eta_{1}$ and $\eta_{2}$, and, therefore, easier to interpret.  Among other things, using the renormalized average densities reduces the effect of the differences in initial conditions seen in Fig.~\ref{fig:Etas_t}.  For instance, it is easy to see in Fig.~\ref{fig:Eta1_Eta2} both the symmetry in the relative dynamics of the two modes and the suppression of the spreading of the condensate wave function with increasing $q$.  Indeed, the harder the trap, the closer $\bar{\eta}_{1}$ and $\bar{\eta}_{2}$ get to 1.

More importantly, the comparison of the dynamics of $\eta_{\alpha}$ with of $\bar{\eta}_{\alpha}$ reveals an interesting fact about the dynamics of our model.  The results for the renormalized average densities are clearly better described by our model than the evolution of $\eta_{1}$ and $\eta_{2}$.  Surprisingly, the same can also be said for the standard reduced-dimension predictions, for there is a dramatic improvement in the agreement between the 3D exact dynamics and the quasi-1D model.  In fact, the standard reduced-dimension approximation failed in all the cases shown in Fig.~\ref{fig:Etas_t}.  Figures~\ref{fig:Eta1_Eta2}(a--c), on the other hand, show that the evolution of $\bar\eta_{1}$ and $\bar\eta_{2}$ can be well described by the quasi-1D GP equation, especially for smaller atom numbers.  These results are very revealing, for they clearly indicate that Eqs.~(\ref{vphi01}) and (\ref{vphi02}) predict, to some extent, the correct relative density dynamics despite using an imprecise initial condition, which is ultimately related to underestimating the transverse profile of the condensate.

Finally, we introduce the renormalized density overlap, $\mathcal{F}_{\eta}$, as the ratio of $\eta_{12}$ to the geometric mean of the single-mode average densities per atom, i.e.,
\begin{equation}
	\mathcal{F}_{\eta}=\frac{\eta_{12}}{\sqrt{\eta_{1}\eta_{2}}}.\label{F_eta}
\end{equation}
This quantity is equal to zero if there is no overlap between the mean-field distributions and is equal to one when they completely overlap, thus working analogously to the fidelity of quantum states.  Figures~\ref{fig:Eta1_Eta2}(d--f) show the time evolution of Eq.~(\ref{F_eta}) for $N=500$, 1\,000, and 2\,000 and the three different traps, $q=2, 4$ and $10$.  This figure nicely shows the time scale on which the condensate density is constant, which increases as the trap gets harder, as expected.  During this time interval, the two-mode dynamics simply correspond to the accumulation of a position-dependent phase-shift.  In addition, the agreement between our model and the 3D numerics is very good for the entire integration time.  This result enforces our previous conclusion that Eqs.~(\ref{vphi01}) and (\ref{vphi02}) predict to some extent the correct relative density dynamics despite the use of an incorrect initial condition.



\section{Optimal reduced-dimension equations for the two-mode evolution \label{sec:optimal_reduced_dimension_evolution_equations}} 

Having successfully obtained a perturbative description of the time-dependent mean fields of the two-mode BEC, we turn now to the question of how to decouple the tightly confined dimensions from the mean-field evolution.

For a single-mode BEC, we showed in Ref.~\cite{tacla11} that, in the stationary case, such a decoupling can be formally achieved by performing a perturbative Schmidt decomposition of the ground-state condensate wave function, $\psi({\bm \rho},{\bm r})$, between the transverse and the longitudinal directions.   In this Schmidt perturbation theory, we seek a solution to the time-independent, single-mode GP equation in the form
\begin{eqnarray}
\label{psiSchmidt}
  \psi({\bm \rho},{\bm r}) = \sqrt{\lambda_0} \chi_0({\bm \rho}) \phi_0({\bm r}) + \epsilon \sqrt{\lambda_1} \chi_1({\bm \rho}) \phi_1({\bm r}) + O(\epsilon^2),
\end{eqnarray}
where $\chi_{n}({\bm \rho})$ and $\phi_{n}({\bm r})$ are the orthonormal Schmidt functions in the transverse and longitudinal directions and the $\sqrt{\lambda_n}$'s are the (nonnegative) Schmidt coefficients (the squares $\lambda_n$ are the eigenvalues of the marginal transverse and longitudinal density matrices)~\cite{nielsenchuang}.  Note that we treat the Schmidt decomposition formally as a power-series expansion in the perturbation parameter $\epsilon$.  Thus the dominant term in Eq.~(\ref{psiSchmidt}) gives the best product approximation to the exact condensate wave function~\cite{lockhart02}, whereas the remaining terms introduce nonseparable corrections and, hence, describe entanglement between the transverse and longitudinal degrees of freedom.  Because in the perturbative regime the spatial entanglement is small~\cite{tacla11}, one can neglect the nonseparable terms and describe the condensate by a product wave function.  In this way, the transverse and longitudinal directions are decoupled and the dominant longitudinal Schmidt function gives the optimal reduced-dimension description for the condensate mean fields.  Below we extend this formalism to the time-dependent mean fields of the two-component BEC, by showing that the instantaneous perturbative Schmidt decomposition can be obtained from the relative-state decomposition~(\ref{timedependentpsiRSD}).

In the single-mode case, the mean field~(\ref{psiSchmidt}) has the form of a two-qubit state to first order in $\epsilon$.  In the two-mode case, however, the single-particle condensate state has the following form
\begin{align}
\label{twomodestate}
	\langle {\bm \rho},{\bm r} |\uppsi_{12}(t)\rangle = \left[\psi^{1}({\bm \rho},{\bm r},t) |1\rangle + \psi^{2}({\bm \rho},{\bm r},t) |2\rangle\right]/\sqrt{2},
\end{align}
which corresponds to a tripartite state, for it has an additional qubit degree of freedom, namely, the two possible hyperfine configurations.  Because the Schmidt decomposition only applies to bipartite systems, the decomposition of tripartite states is not unique and depends on the specific way one chooses to partition the system.

From all possible bipartitions, the one that is physically relevant for understanding the reduced-dimension structure of the mean field of a highly anisotropic condensate is the one that separates the transverse degrees of freedom from the longitudinal and hyperfine (internal) degrees of freedom. This partition relies on the physical difference between the high energy scale of the tightly confined dimensions and the much lower energy scale of the longitudinal and hyperfine degrees of freedom.

The procedure we describe next is very similar to the one discussed in Appendix~B of Ref.~\cite{tacla11}.  We start by pointing out that the two-mode, single-particle state~(\ref{twomodestate}) can be written in the form of the spinor
\begin{align}
	\vec{\uppsi}_{12} = \frac{1}{\sqrt{2}}\colvec{\psi^{1}}{\psi^{2}},\label{psi_spinor}
\end{align}
which represents a vector in the two-dimensional hyperfine space, whose components are the condensate wave functions
\begin{equation}
\psi^{\alpha}= e^{-iE_0 t/\hbar} \left( \xi_0\upvarphi^{\alpha}_0 + \epsilon\sum_{n=1}^\infty \xi_n \upvarphi^{\alpha}_n\right),
\end{equation}
and which satisfies the normalization condition,
\begin{equation}
	\langle\vec{\uppsi}_{12}|\vec{\uppsi}_{12}\rangle =
\frac{1}{2}\left(\langle \psi^1|\psi^1\rangle+\langle\psi^2|\psi^2\rangle\right)
=1 + O(\epsilon^2).
\end{equation}
For simplicity, from here on we remove the initial-condition terms from our equations by assuming that $\upvarphi_n^\alpha(0)=\tilde\upvarphi_n^\alpha(0)$, which implies that $\upvarphi_n^\alpha=\tilde\upvarphi_n^\alpha$ holds at all times. As we showed in Sec.~\ref{sub:source_terms} and in Fig.~\ref{fig:OverlapQuinticNoSource1000}, for the initial conditions we use in this paper, these initial-condition terms make virtually no difference. For other initial conditions, however, these terms might need to be included, and they could be added trivially to the derivation below.

We now move toward a Schmidt decomposition by separating out the transverse degrees of freedom from the other ones in Eq.~(\ref{psi_spinor}),
\begin{align}
	\vec{\uppsi}_{12} = \upzeta_0\vec{\upvarphi}_0 + \epsilon\upzeta_1 \vec{\upvarphi}_1,\label{psi_spinor_bipartite}
\end{align}
where
\begin{align}
	\upzeta_0 = \xi_0 e^{-i E_0t/\hbar},\qquad
	\upzeta_1 = -e^{-i E_0t/\hbar} \sum_{n=1}^\infty \xi_n \frac{\langle \xi_{n} | \xi_{0}^{3} \rangle}{E_n-E_0},
\end{align}
and
\begin{align}
	\vec{\upvarphi}_0 = \frac{1}{\sqrt{2}}\colvec{\upvarphi_0^1}{\upvarphi_0^2},\qquad
	\vec{\upvarphi}_1 = \frac{1}{\sqrt{2}}
\colvec{\bigl(\tilde{g}_{11} |\upvarphi^{1}_{0}|^{2} + \tilde{g}_{12}|\upvarphi^{2}_{0}|^{2}\bigr)\upvarphi^{1}_{0}}
{\bigl(\tilde{g}_{22} |\upvarphi^{2}_{0}|^{2} + \tilde{g}_{12} |\upvarphi^{1}_{0}|^{2}\bigr)\upvarphi^{2}_{0}}.
\end{align}

We can formulate the two-term decomposition~(\ref{psi_spinor_bipartite}) because, as in the static case~\cite{tacla11}, we can segregate the $n$ dependence of the $n>1$ corrections in a transverse piece, leaving a longitudinal-hyperfine piece that has no $n$ dependence.  The resulting decomposition is not yet in Schmidt form, for the functions are not orthonormal, but we can easily transform it to Schmidt form.  We begin by noting that
\begin{align}
	\langle \upzeta_0|\upzeta_{0}\rangle &= 1,\\
	\langle \upzeta_0|\upzeta_{1}\rangle &= 0,\\	
	\langle \upzeta_1|\upzeta_1\rangle &= \sum_{n=1}^\infty  \frac{\langle \xi_{n} | \xi_{0}^{3} \rangle^2}{(E_n-E_0)^2},\\
	\langle \vec{\upvarphi}_0|\vec{\upvarphi}_0\rangle &= 1,\\
	\langle \vec{\upvarphi}_0|\vec{\upvarphi}_1\rangle &= \frac{1}{2}\left(\tilde{g}_{11}\eta_L^1 + \tilde{g}_{22}\eta_L^2 + 2\tilde{g}_{12}\eta_L^{12}\right),\\
	\langle \vec{\upvarphi}_1|\vec{\upvarphi}_1\rangle &= \frac{1}{2}\left(\tilde{g}_{11}^2\eta_{L6}^1 + \tilde{g}_{22}^2\eta_{L6}^2 + (2\tilde{g}_{11}+\tilde{g}_{12})\tilde{g}_{12}\eta_L^{112}+ (2\tilde{g}_{22}+\tilde{g}_{12})\tilde{g}_{12}\eta_L^{221}\right).
\end{align}
We now write
\begin{align}
	\vec{\uppsi}_{12} &= \big(\upzeta_0+\epsilon\langle \vec{\upvarphi}_0|\vec{\upvarphi}_1\rangle\upzeta_1\big)\vec{\upvarphi}_0 + \epsilon\upzeta_1 \big(\vec{\upvarphi}_1-\langle \vec{\upvarphi}_0|\vec{\upvarphi}_1\rangle\vec{\upvarphi}_0\big),
\end{align}
which inspires us to define the following functions,
\begin{align}
	\upchi_0 &= \upzeta_0+\epsilon\langle \vec{\upvarphi}_0|\vec{\upvarphi}_1\rangle\upzeta_1,\qquad \upchi_1 = \frac{\upzeta_1}{\sqrt{\langle \upzeta_1|\upzeta_1\rangle}},\label{upchi0upchi1}\\
	\vec{\upphi}_0 &= \vec{\upvarphi}_0,\qquad \vec{\upphi}_1 = \frac{\vec{\upvarphi}_1-\langle \vec{\upvarphi}_0|\vec{\upvarphi}_1\rangle\vec{\upvarphi}_0}{\sqrt{\langle \vec{\upvarphi}_1|\vec{\upvarphi}_1\rangle - |\langle \vec{\upvarphi}_0|\vec{\upvarphi}_1\rangle|^{2}}}.\label{upphi0upphi1}
\end{align}
At the order we are working, $\upchi_0$ and $\upchi_1$ are orthonormal, whereas $\vec{\upphi}_0$ and $\vec{\upphi}_1$ are orthonormal by construction.   Thus Eqs.~(\ref{upchi0upchi1}) and (\ref{upphi0upphi1}) form the perturbative Schmidt basis of the two-mode condensate state~(\ref{psi_spinor_bipartite}).  Therefore, we write
\begin{align}
	\vec{\uppsi}_{12} = \sqrt{\uplambda}_0\upchi_0\vec{\upphi}_0 + \epsilon\, \sqrt{\uplambda}_1\upchi_1\vec{\upphi}_1,\label{psi_spinor_schmidt}
\end{align}
where the squared Schmidt coefficients are given by
\begin{align}
	\uplambda_0 = 1,\qquad
	\uplambda_1 = \langle \upzeta_1|\upzeta_1\rangle\left( \langle \vec{\upvarphi}_1|\vec{\upvarphi}_1\rangle - |\langle \vec{\upvarphi}_0|\vec{\upvarphi}_1\rangle|^{2} \right).\label{uplambda}
\end{align}

As in Ref.~\cite{tacla11}, we use Wootters's concurrence~\cite{wootters98} to quantify the bipartite entanglement of the state~(\ref{psi_spinor_schmidt}).  The concurrence of a pure state of two qubits, $|\Psi_{AB}\rangle$, varies smoothly from 0 for product states to 1 for maximally entangled states.  From its definition, $C = |\langle \Psi_{AB}^* | \sigma_y\otimes\sigma_y | \Psi_{AB}^* \rangle|$, in terms of the Pauli matrix $\sigma_y$ and the complex conjugate of $|\Psi_{AB}\rangle$, it is easy to show that the concurrence for the perturbative condensate single-particle state~(\ref{psi_spinor_schmidt}) is given by
\begin{equation}
	C = 2\sqrt{\uplambda_0 \uplambda_1} = 2\langle \upzeta_1|\upzeta_1\rangle\left( \langle \vec{\upvarphi}_1|\vec{\upvarphi}_1\rangle - |\langle \vec{\upvarphi}_0|\vec{\upvarphi}_1\rangle|^{2} \right),
\end{equation}
where we use the perturbative coefficients $\uplambda_0$ and $\uplambda_1$ as given by Eq.~(\ref{uplambda}).

In the case of a cigar-shaped BEC confined in a transverse harmonic potential,
\begin{equation}
	\langle \upzeta_1|\upzeta_1\rangle = \left( \frac{\eta_T }{2\hbar\omega_T} \right)^2 {\rm Li}_2(1/4),
\end{equation}
where we use the polylogarithm function ${\rm Li}_{s}(z)\equiv\sum_{n=1}^\infty z^n/n^s$.  Thus, it follows that the time-dependent concurrence can be written as
\begin{align}
	C = &\left(\frac{\eta_T }{2\hbar\omega_T} \right)^2 {\rm Li}_2(1/4) \Bigg[ \frac{\tilde{g}_{11}^2}{2} \Big( \eta_{L6}^1 - (\eta_L^1)^2 \Big) + \frac{\tilde{g}_{22}^2}{2} \Big( \eta_{L6}^2 - (\eta_L^2)^2 \Big) \nonumber\\
	&+ \tilde{g}_{12}^2 \Big( \eta_{L}^{112} + \eta_{L}^{221} - 2 (\eta_L^{12})^2 \Big) + \frac{1}{2}\big(\tilde{g}_{11}^2\eta_{L6}^1 + \tilde{g}_{22}^2\eta_{L6}^2 - 2\tilde{g}_{11}\tilde{g}_{22}\eta_L^1\eta_L^2\big)\nonumber\\
	&+ 2\tilde{g}_{12} \big(\tilde{g}_{11}\eta_{L}^{112} + \tilde{g}_{22}\eta_{L}^{221} - \tilde{g}_{11}\eta_L^1\eta_L^{12} - \tilde{g}_{22}\eta_L^2\eta_L^{12}\big) \Bigg].\label{concurrence_spinor}
\end{align}

\begin{figure}[!htbp]
	\centering
		\includegraphics[width=\textwidth]{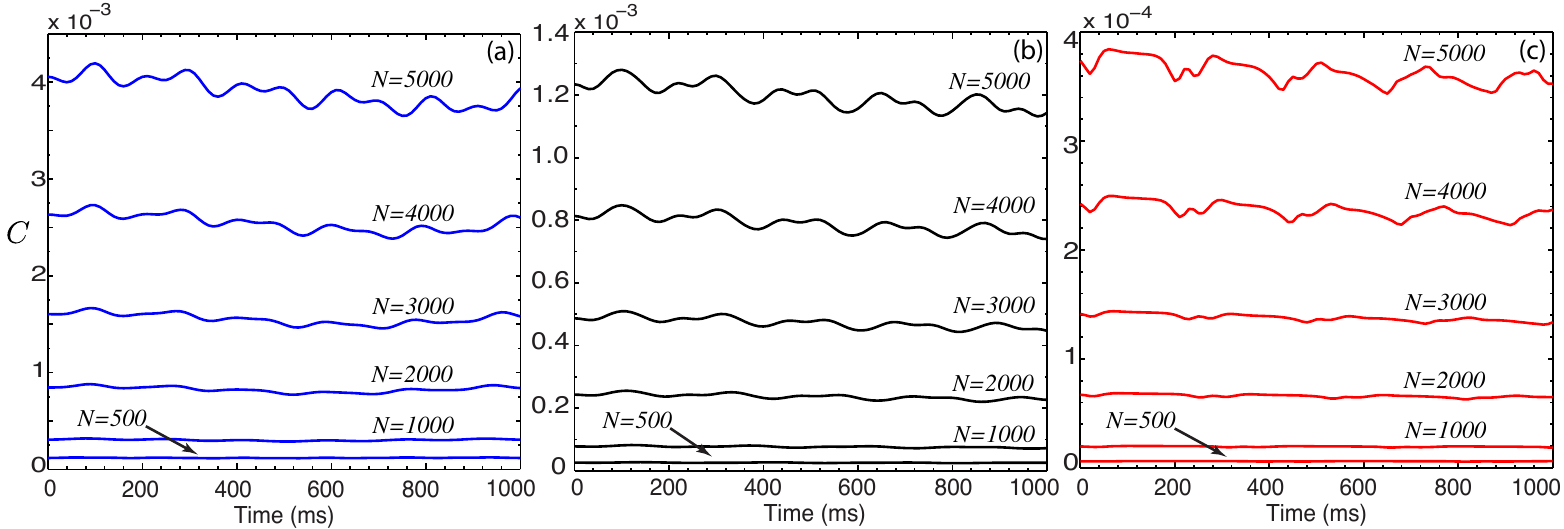}
	\caption{Concurrence~(\ref{concurrence_spinor}) of the two-mode condensate state~(\ref{psi_spinor_schmidt}) as a function of time and atom number, calculated using the parameters given in section~\ref{sec:two_mode_dynamics_of_a_cigar_shaped_87_rb_condensate} for (a)~$q=2$, (b)~$q=4$, and (c)~$q=10$.  The entanglement remains remarkably small, indicating that the condensate is well approximated by a single Schmidt term for atom numbers as high as 5\,000. Moreover, the entanglement between the transverse dimensions and the longitudinal-hyperfine degrees of freedom decreases as the potential becomes more homogeneous. Note that the concurrence for $q=10$ has the lowest values.  To a very good approximation the nonseparable corrections are indeed negligible.}
	\label{fig:ConcurrenceTwoMode3}
\end{figure}

We display the concurrence~(\ref{concurrence_spinor}) in figures~\ref{fig:ConcurrenceTwoMode3}(a)--(c) as a function of time and atom number, calculated using the parameters given in section~\ref{sec:two_mode_dynamics_of_a_cigar_shaped_87_rb_condensate} for $q=2$, 4, and 10.  In all cases, the concurrence oscillates in time as well as increases as $N$ gets larger. More importantly, we can verify that the entanglement between the transverse and the remaining degrees of freedom is quite small for all cases, even for larger atom numbers.  Moreover, it also decreases as the trap gets harder.  These results indicate that the two-mode condensate state~(\ref{psi_spinor_schmidt}) is well described by the product state corresponding to the first Schmidt term.  As we discussed before, the dominant term in the Schmidt decomposition corresponds to the optimal product-state approximation to state~(\ref{twomodestate}).  This result not only confirms that the tightly confined dimensions can indeed be decoupled from the problem, but it naturally shows the optimal way to do it.


\section{Conclusion} 
\label{sec:conclusion}

We have presented a detailed theoretical analysis of the mean-field dynamics and the reduced-dimension character of two-mode BECs in highly anisotropic traps.  These systems are promising candidates for the implementation of high-precision nonlinear interferometers, and our analysis shows the feasibility of a recent experimental proposal of a Ramsey interferometry scheme~\cite{boixo08b}.

In spite of the strong anisotropy of the condensate, our results also show that the effects of the nonlinear scattering interaction on the transverse degrees of freedom of the gas cannot be in general neglected.  This brings into question the accuracy of the standard reduced-dimension (quasi-1D) approximation, both spatially and temporally.   We have formulated a model that takes into account such 3D-induced effects by means of perturbative techniques.  Because of the reshaping of the condensate in the transverse direction(s), additional effective three-body, attractive, intra- and inter-mode interactions have to be included in the reduced-dimension equations of the longitudinal dimension(s).  Moreover, we have shown that these equations in fact provide the optimal reduced-dimension evolution for the two-mode condensate in general quasi-1D and quasi-2D geometries.  We have demonstrated this result by developing a Schmidt decomposition of the two-mode condensate single-particle wave function, which separates out the transverse from the longitudinal-internal degrees of freedom.  Using this formalism, we have verified that the entanglement between the transverse and the remaining degrees of freedom is quite small, even for larger atom numbers.  In other words, this result shows that the two-mode condensate single-particle state is well described by the product state corresponding to the dominant term in the Schmidt decomposition.  This corresponds to the optimal product-state approximation~\cite{lockhart02}.

Finally, as far as the interferometry process goes, our perturbative model takes into account the spatial differentiation of the wave functions of the two modes and gives a significantly more accurate analytical account of the fringe signal than does the standard reduced-dimension approximation.  We point out that this theoretical model is not limited to the study of interference effects, but can also be applied to the study of more general dynamics and different phenomena, such as soliton propagation.


\ack

We thank A.~Shaji for discussions at the beginning of this project.  This work was supported in part by National Science Foundation Grant Nos.~PHY-0903953 and~PHY-1005540.

\appendix

\section{Adiabatic elimination of $\upvarphi_n^{\alpha}$} 
\label{sec:adiabatic_elimination_of_varphi_n}

We start by integrating Eq.~(\ref{vphinsol}) by parts, which lets us write the following relation for $\upvarphi^{\alpha}_n(t)$, $n\ge1$:
\begin{align}
\label{vphinparts}
	\upvarphi_n^{\alpha}(t) &- \upvarphi_n^{\alpha}(0) e^{-i (E_n-E_0)t/\hbar}\nonumber\\
&= -\frac{\langle \xi_{n} | \xi_{0}^{3} \rangle}{E_n-E_0} e^{-i (E_n-E_0)(t-s)/\hbar}
\Big( \tilde{g}_{\alpha \alpha} |\upvarphi^{\alpha}_{0}(s)|^{2}
+ \tilde{g}_{\alpha \beta} |\upvarphi^{\beta}_{0}(s)|^{2} \Big) \upvarphi^{\alpha}_{0}(s)\Big|_0^t + O(\epsilon).
\end{align}
Here the first term on the right-hand side arises from the straightforward integration of the rapidly oscillating exponential in Eq.~(\ref{vphinsol}), whereas the (neglected) remaining term is of higher order in $\epsilon$ because it involves time derivatives of $\upvarphi^{\alpha}_0$ and $\upvarphi_0^{\alpha*}$.  We assume from now on, consistent with our perturbation theory, that these higher-order terms in Eq.~(\ref{vphinparts}) can be neglected.  Under this approximation, we write
\begin{align}
\label{vphinapprox}
\upvarphi_n^{\alpha}(t) \simeq \tilde{\upvarphi}_n^{\alpha}(t) + \tilde{I}_n^{\alpha}(t),
\end{align}
where
\begin{align}
\label{vphinadiabatic}
\tilde{\upvarphi}_n^{\alpha}(t) &= - \frac{\langle \xi_{n} | \xi_{0}^{3} \rangle}{E_n-E_0} \left( \tilde{g}_{\alpha \alpha} |\upvarphi^{\alpha}_{0}(t)|^{2} + \tilde{g}_{\alpha \beta} |\upvarphi^{\beta}_{0}(t)|^{2} \right) \upvarphi^{\alpha}_{0}(t)
\end{align}
corresponds to the adiabatic following of $|\upvarphi^{\alpha}_{0}|^{2}\upvarphi^{\alpha}_{0}$ and $|\upvarphi^{\beta}_{0}|^{2}\upvarphi^{\alpha}_{0}$, which is the dominant dynamical effect described by Eq.~(\ref{vphinparts}).  The remaining lowest-order term,
\begin{equation}
	\tilde{I}_n^{\alpha}(t) = \big( \upvarphi_n^{\alpha}(0) - \tilde{\upvarphi}_n^{\alpha}(0)\big) e^{-i (E_n-E_0)t/\hbar},\label{Inalpha}
\end{equation}
is associated with the deviation of $\upvarphi_n^{\alpha}(0)$, the projection of the initial condensate wave function onto $\xi_n$, from its adiabatic variant $\tilde{\upvarphi}_n^{\alpha}(0)$.

We can now use the adiabatic approximation~(\ref{vphinapprox}) to eliminate $\upvarphi^{\alpha}_n(t)$, $n\ge1$, from Eq.~(\ref{vphi0expansion}), thus obtaining a more tractable evolution equation for $\upvarphi^{\alpha}_0$ than Eq.~(\ref{vphi0}):
\begin{align}
\label{vphi0elimination}
	i\hbar \dot{\upvarphi}^{\alpha}_{0}=\epsilon&\left( H_L + \tilde{g}_{\alpha \alpha} \eta_T |\upvarphi^{\alpha}_{0}|^{2} + \tilde{g}_{\alpha \beta} \eta_T |\upvarphi^{\beta}_{0}|^{2} \right)\upvarphi^{\alpha}_{0}\nonumber\\
	& - \epsilon^{2} \UpsilonT \left( 3\tilde{g}_{\alpha \alpha} |\upvarphi_0^{\alpha}|^{2} + \tilde{g}_{\alpha \beta} |\upvarphi_0^{\beta}|^{2} \right)\left( \tilde{g}_{\alpha \alpha} |\upvarphi^{\alpha}_{0}|^{2} + \tilde{g}_{\alpha \beta} |\upvarphi^{\beta}_{0}|^{2} \right) \upvarphi^{\alpha}_{0}\nonumber\\
	& - 2\epsilon^{2} \UpsilonT \tilde{g}_{\alpha \beta} \upvarphi_0^{\alpha} |\upvarphi^{\beta}_{0}|^{2} \left( \tilde{g}_{\beta \beta} |\upvarphi^{\beta}_{0}|^{2} + \tilde{g}_{\beta \alpha} |\upvarphi^{\alpha}_{0}|^{2} \right) + \epsilon^{2} \tilde{I}_0^{\alpha}.
\end{align}
Here
\begin{align}
	\tilde{I}_0^{\alpha} =& \sum_{n=1}^{\infty}\langle \xi_{0}^{3} | \xi_{n} \rangle \nonumber\\
&\times\Big[  \Big( 2\tilde{g}_{\alpha \alpha} |\upvarphi_0^{\alpha}|^{2} + \tilde{g}_{\alpha \beta} |\upvarphi_0^{\beta}|^{2} \Big) \tilde{I}_n^{\alpha} +  \tilde{g}_{\alpha \alpha} (\upvarphi_0^{\alpha})^{2} \tilde{I}_n^{\alpha*} + \tilde{g}_{\alpha \beta} \upvarphi_0^{\alpha} \Big( \upvarphi_0^{\beta}\tilde{I}_n^{\beta*} + \upvarphi_0^{\beta*}\tilde{I}_n^{\beta} \Big) \Big],
\end{align}
and the coupling parameter $\UpsilonT$ is defined by Eq.~(\ref{UpsilonT}).

After reorganizing the second-order terms in Eq.~(\ref{vphi0elimination}), we obtain the final form of the effective, coupled, two-mode evolution equation for the dominant longitudinal wave function $\upvarphi_0^{\alpha}$,
\begin{align}
\label{vphi0final}
	i\hbar \dot{\upvarphi}^{\alpha}_{0}
	&= \epsilon \left( H_L + \tilde{g}_{\alpha \alpha} \eta_T |\upvarphi^{\alpha}_{0}|^{2} + \tilde{g}_{\alpha \beta} \eta_T |\upvarphi^{\beta}_{0}|^{2} \right)\upvarphi^{\alpha}_{0} \nonumber\\
	&\quad - \epsilon^{2} \UpsilonT \Big( 3 \tilde{g}_{\alpha \alpha}^2 |\upvarphi^{\alpha}_{0}|^{4} + (4 \tilde{g}_{\alpha \alpha} + 2\tilde{g}_{\beta \alpha})\tilde{g}_{\alpha \beta} |\upvarphi_0^{\alpha}|^{2}|\upvarphi^{\beta}_{0}|^{2}+
(\tilde{g}_{\alpha \beta} + 2\tilde{g}_{\beta \beta}) \tilde{g}_{\alpha \beta}|\upvarphi_0^{\beta}|^{4}\Big)\upvarphi^{\alpha}_{0}\nonumber\\
&\quad + \epsilon^{2}  \tilde{I}^{\alpha}_0,
\end{align}
where
\begin{align}
	\tilde{I}_0^{\alpha} &= \left( 2\tilde{g}_{\alpha \alpha} |\upvarphi_0^{\alpha}|^{2} + \tilde{g}_{\alpha \beta} |\upvarphi_0^{\beta}|^{2} \right) (\Phi^{\alpha} - \tilde{\Phi}^{\alpha\beta}) +  \tilde{g}_{\alpha \alpha} (\upvarphi_0^{\alpha})^{2} (\Phi^{\alpha*} - \tilde{\Phi}^{\alpha\beta*}) \nonumber\\
	&\quad + \tilde{g}_{\alpha \beta} \upvarphi_0^{\alpha} \left[ \upvarphi_0^{\beta}(\Phi^{\beta*} - \tilde{\Phi}^{\beta\alpha*}) + \upvarphi_0^{\beta*}(\Phi^{\beta} - \tilde{\Phi}^{\beta\alpha}) \right].\label{I0alpha}
\end{align}
Here we introduce the functions
\begin{align}
	\Phi^{\alpha}(t) &= \sum_{n=1}^{\infty} \langle \xi_{n} | \xi_{0}^{3} \rangle \upvarphi_n^{\alpha}(0) e^{-i (E_n-E_0)t/\hbar}, \label{Phia}\\		
	\tilde{\Phi}^{\alpha\beta}(t) &= -\left( \tilde{g}_{\alpha \alpha} |\upvarphi^{\alpha}_{0}(0)|^{2} + \tilde{g}_{\alpha \beta} |\upvarphi^{\beta}_{0}(0)|^{2} \right)\upvarphi^{\alpha}_{0}(0)\upsilonT(t),\label{Phiab}
\end{align}
where
\begin{equation}
	\upsilonT(t) = \sum_{n=1}^{\infty} \frac{\langle \xi_{n} | \xi_{0}^{3} \rangle^2}{E_n-E_0} e^{-i (E_n-E_0)t/\hbar}.\label{upsilonT}
\end{equation}
is a response function that satisfies $\upsilonT(0)=\UpsilonT$ and $\dot\upupsilon_T(t)=\eta_T^2 G_T(t)$.

The importance of the source term $\tilde{I}_0^{\alpha}$ in determining the dynamical behavior of $\upvarphi^{\alpha}_0$ essentially depends on the difference between $\upvarphi_n^{\alpha}(0)$ and $\tilde{\upvarphi}_n^{\alpha}(0)$.  As shown by Eqs.~(\ref{Phia}) and (\ref{Phiab}), $\tilde{I}_0^{\alpha}$ is a rapidly varying function in comparison to the other terms in Eq.~(\ref{vphi0final}).  In the case of a transverse harmonic confinement, we can write $\upsilonT(t)$ in the closed form
\begin{equation}
\label{upsilonTharmonic}
\upsilonT(t)=
\begin{cases}
\displaystyle{\vphantom{\Bigg(\biggr)}\, -\frac{\eta_T^2}{\hbar\omega_T} \text{ln}\left[\frac{1}{2} \left(1+\sqrt{1-\frac{1}{4}e^{-2i\omega_T t}}\right)\right] ,}&\mbox{$D=1$ (pancake),}\\
\displaystyle{\vphantom{\Bigg(\biggr)}-\frac{\eta_T^2}{2\hbar\omega_T}\, \text{ln\!}\left(1-\frac{1}{4} e^{-2i\omega_T t}\right) ,}&\mbox{$D = 2$ (cigar).}
\end{cases}
\end{equation}

\section{Single-mode dynamics within the adiabatic elimination of $\upvarphi_n^{\alpha}$} 
\label{sub:single_mode_dynamics_within_the_adiabatic_elimination_of_upvarphi_n_alpha}

The dynamical equations for a single-mode BEC can be obtained trivially from Eqs.~(\ref{vphinapprox}) and (\ref{vphi0final}).  By simply setting $\tilde{g}_{\alpha \beta}=0$ and dropping the superfluous index $\alpha$, we get
\begin{align}
		\upvarphi_n &= - \frac{\langle \xi_{n} | \xi_{0}^{3} \rangle}{E_n-E_0} \tilde{g} |\upvarphi_{0}|^{2} \upvarphi_{0} + \tilde{I}_n
=\tilde\upvarphi_n+\tilde I_n,\quad n\ge1,\label{vphinsinglemode}\\
	i\hbar \dot{\upvarphi}_{0} &= \epsilon \left( H_L + \tilde{g} \eta_T |\upvarphi_{0}|^{2} - 3\epsilon \tilde{g}^2 \UpsilonT |\upvarphi_{0}|^{4} \right)\upvarphi_{0} + \epsilon^2 \tilde{I}_0,\label{vphi0singlemode}
\end{align}
where
\begin{align}
	\tilde{I}_n &= \big( \upvarphi_n(0) - \tilde{\upvarphi}_n(0)\big) e^{-i (E_n-E_0)t/\hbar}, \\
	\tilde{I}_0 &= 2\tilde{g} |\upvarphi_0|^{2} (\Phi - \tilde{\Phi}) +  \tilde{g} \upvarphi_0^{2} (\Phi^{*} - \tilde{\Phi}^{*}),
\end{align}
and
\begin{align}
	\Phi(t) &= \sum_{n=1}^{\infty} \langle \xi_{n} | \xi_{0}^{3} \rangle \upvarphi_n(0) e^{-i (E_n-E_0)t/\hbar},\\
	\tilde{\Phi}(t) &= -\tilde{g} |\upvarphi_{0}(0)|^{2} \upvarphi_{0}(0)\upsilonT(t).
\end{align}

Equations~(\ref{vphinsinglemode}) and (\ref{vphi0singlemode}) are the time-dependent versions of the stationary single-mode equations derived in Ref.~\cite{tacla11}, but with the additional source terms $\tilde{I}_n$ and $\tilde{I}_0$.  In fact, in the stationary case, $\upvarphi_n = \tilde{\upvarphi}_n$, and hence $\Phi = \tilde{\Phi}$. As a result, both $\tilde{I}_n$ and $\tilde{I}_0$ vanish, and the equations reduce to those of Ref.~\cite{tacla11}.  Note that relative to a GP equation, the longitudinal equation~(\ref{vphi0singlemode}) has an additional quintic term, which acts as an effective three-body, attractive interaction among the atoms regardless of the sign of $g$.  This attractive interaction is mediated by the changes in the transverse wave function, as evidenced by the appearance of the (nonnegative) coupling parameter $\UpsilonT$ in the coupling strength $3g^2\UpsilonT$.

The coupling constants in Eq.~(\ref{vphi0singlemode}) can be calculated explicitly for a transverse harmonic potential.  It follows from Eqs.~(\ref{overlap1D}), (\ref{overlap2D}), and (\ref{UpsilonT}) that
\begin{equation}
\label{Upsilon_harmonic}
\UpsilonT=
\begin{cases}
\displaystyle{\vphantom{\Bigg(\biggr)}\frac{\eta_T^2}{\hbar\omega_T}\ln(8-4\sqrt{3}),}&\mbox{$D=1$ (pancake),}\\
\displaystyle{\vphantom{\Bigg(\biggr)}\frac{\eta_T^2}{2\hbar\omega_T}\ln\frac{4}{3},}&\mbox{$D = 2$ (cigar).}
\end{cases}
\end{equation}
As a result, the coupling constants for a cigar-shaped trap ($d=1$) are $g \eta_T = 2 \hbar\omega_T a$ and $3g^2\UpsilonT = 6\hbar\omega_T a^2\ln(4/3)$, whereas for a quasi-2D pancake ($d=2$), we obtain $g \eta_T = 2\sqrt{2\pi}\hbar\omega_T \rho_0 a$ and $3g^2\UpsilonT = 24\pi\hbar\omega_T \rho_0^2a^2\ln(8-4\sqrt{3})$.

It is worth pointing out that such a self-focusing interaction has been used to study the propagation of solitons in single-mode, attractive, quasi-1D condensates, trapped by an infinitely long cylindrical harmonic potential~\cite{muryshev02,sinha06,khaykovich06}.  In contrast to using a relative-state decomposition of the three-dimensional condensate wave function, those studies use a Taylor expansion of a local, transverse chemical potential about the maximum density of the condensate to derive the additional quintic nonlinearity.  This leads to an equation without the source term $\tilde{I}_0$ and with a coupling constant that differs from $3 g^2 \UpsilonT$ (for $d = 1$) by a factor of 4.  This factor corresponds to the difference between using the maximum of the Gaussian ground-state probability distribution and the average $\eta_T$ of the distribution over itself.




\begin{thebibliography}{100}
\bibitem{note1}
	This approximation holds as long as the Rabi-flopping time scale for the atomic transition is smaller than the time scale for the transverse BEC dynamics (the longitudinal BEC dynamics occurs on an even longer time scale).  For the \rbd~BECs we consider here, the Rabi-flopping time scale is typically less than a millisecond, so these conditions hold for the trapping potentials we consider here.
\bibitem{note}
	In our simulations we assume that one can neglect the evolution of
	the spatial modes during the Ramsey pulses.  According to Ref.~\cite
	{mertes07a,anderson09,matthews98a}, the two hyperfine levels of
	$^{87}$Rb can be coupled by two-photon Rabi pulses whose duration is
	less than 1 ms, which is much faster than the characteristic
	time for the condensate to change its shape, thereby justifying such an
	approximation.
\bibitem{units}
	Plots that are in harmonic trap units have (i)~transverse (longitudinal) lengths in units of $\rho_0$ ($z_0$ for $q=2$), (ii)~transverse (longitudinal) marginal distributions in units of $1/\rho_0^{2}$ ($1/z_0$ for $q=2$), and (iii)~density in units of $1/(\rho_0^{2}z_0)$.
\end{thebibliography}
\end{document}